\documentclass[a4paper,11pt]{article}
\pdfoutput=1 

\usepackage{jheppub} 

\usepackage[T1]{fontenc} 

\newcommand{\bx}{{\bf x}}
\newcommand{\bk}{{\bf k}}
\newcommand{\omegat}{\tilde{\omega}}
\newcommand{\xt}{\tilde{x}}

\def\be   {\begin{equation}}   \def\ee   {\end{equation}}

\title{\boldmath Entanglement and Expansion}

\author[a]{K. Boutivas,}
\author[b,c]{G. Pastras}
\author[a]{and N. Tetradis}

\affiliation[a]{Department of Physics, University of Athens,\\Zographou 157 84, Greece}
\affiliation[b]{Institute of Nuclear and Particle Physics, NCSR `Demokritos',\\ Aghia Paraskevi 15310, Greece}
\affiliation[c]{Laboratory for Manufacturing Systems and Automation,\\ Department of Mechanical Engineering and Aeronautics,\\ University of Patras, Patra 26110, Greece}

\emailAdd{kboutivas@phys.uoa.gr}
\emailAdd{pastras@lms.mech.upatras.gr}
\emailAdd{ntetrad@phys.uoa.gr}

\abstract{We study the entanglement entropy resulting from tracing out local degrees of freedom of a quantum scalar field in an expanding universe. It is known that when field modes become superhorizon during inflation they evolve to increasingly squeezed states. We argue that this causes the entanglement entropy to grow continuously as successive modes cross the horizon. The resulting entropy is proportional to the total duration of inflation. It is preserved during a subsequent radiation or matter dominated era, and thus it may be relevant for today's universe. We demonstrate explicitly these features in a toy model of a scalar field in 1+1 dimensions.}

\begin{document}
\maketitle
\flushbottom

\section{Introduction}

In inflationary cosmology a fluctuating scalar field can be expressed in terms of 
momentum modes whose mode function involves 
two terms: one with constant magnitude that corresponds to a free field in flat space; 
and one that grows indefinitely with conformal time. 
When the mode wavelength crosses the horizon, the scalar fluctuation gets
dominated by the second term and loses its oscillatory form (it freezes) \cite{physrep}. 
Moreover, the domination of the
growing contribution causes the field and its conjugate momentum to commute.
After horizon crossing, the field can be viewed as a classical stochastic 
field, and its quantum expectation value can be considered as the classical 
stochastic average.

From a quantum mechanical point of view, the modes of the scalar field 
evolve from a simple oscillator ground state to an increasingly squeezed state, 
which ceases to be a minimal uncertainty state \cite{Grishchuk}. 
Even though the two pictures are equivalent, it is assumed that 
the quantum properties of the field are not visible in late-time 
observations which focus on classical local quantities \cite{albrecht,classical1,classical2,classical3,classical4,classical5}.
However, there is a purely quantum non-local phenomenon
that does not have a classical analogue, namely quantum entanglement, 
which cannot be encoded in the classical probability distributions. 
In general, 
the squeezing of canonical modes increases the entanglement between 
local degrees of freedom and is expected to increase the entanglement entropy. 
There is a long history of studies on the connection between 
squeezing and entropy \cite{squeeze1,squeeze2,squeeze3,squeeze4,squeeze5,squeeze6,squeeze7,squeeze8}.

The calculation of the entanglement entropy is based on the division of the overall degrees of freedom to two subsets and the determination 
of the reduced density matrix, which describes one of the two subsets, 
via the tracing out of the degrees of freedom of the other. 
We are interested here in the case that
these degrees of freedom correspond to the interior of a spherical entangling surface.
It is known that, for a scalar field theory in flat space, the resulting entanglement 
entropy is dominated by an area law term \cite{Bombelli:1986rw,srednicki}, 
displaying a profound resemblance to the black hole entropy.
In general, explicit calculations are possible only
for non-interacting or highly symmetric quantum field theories 
\cite{Bombelli:1986rw,srednicki,muller,casini0,casini1,casini2,pimentel,wilczek,korepin,cardy1,cardy2,Kanno:2014lma,Iizuka:2014rua,Kanno:2016qcc,stefan,Lohmayer}.
The holographic approach and the Ryu-Takayanagi proposal
\cite{ryu1,ryu2,review1,review2} 
provide a simpler framework for theories that have a gravitational
dual in the context of the AdS/CFT correspondence \cite{adscft1,adscft2,adscft3}.
The demanding calculation
of the density matrix is replaced by the calculation of the area
of a minimal surface that is anchored at the entangling surface on the boundary of AdS space 
and extends into the bulk. This framework confirms the area law 
in flat space, but also allows the calculation of the 
entanglement entropy for time-dependent backgrounds \cite{tetradisgiataganas,tetradisgiantsos}.

In more recent years, a lot of attention has been attracted to the possible connection 
of quantum entanglement  to gravity. 
The AdS/CFT correspondence suggests that the gravity in the bulk theory may emerge 
as a quantum entropic force due to quantum statistics related to 
entanglement \cite{Lashkari:2013koa}. 
A key feature of entanglement that supports such 
a conjecture \cite{Jacobson:entanglement1,Jacobson:entanglement2} is 
the fact that the leading contribution to the
entanglement entropy in scalar field theory follows an area law \cite{Bombelli:1986rw,srednicki}, 
similarly to the black hole entropy.

As we mentioned above, 
the squeezing of the field modes in an expanding universe has 
a strong effect on quantum entanglement. 
Several studies have
focused mainly on entanglement between different momentum modes,
e.g. subhorizon and superhorizon modes \cite{momentumspace1,momentumspace2}. 
However, a more natural separation of the degrees of freedom would 
be based on the definition of an entangling surface. 
For example, the cosmological horizon naturally divides the degrees of freedom 
to observable and non-observable ones according to a given observer. 
Then, the entanglement entropy could contribute to the cosmological horizon entropy.
Phenomenological implications of the entanglement between coarse-grained configurations
of a quantum field in real space have been studied in 
\cite{colas,vennin1,vennin2,vennin3}.

Studying the entanglement between spatially separated degrees of freedom in an expanding 
universe demands the extension of the techniques of \cite{srednicki} to the case of 
squeezed states. In general the extension of these methods beyond the usual ground state of 
the simple harmonic oscillator presents a high degree of complexity (see e.g. 
\cite{casini0} for a review of entanglement entropy in free QFT). In this work, we adopt 
some elements of \cite{srednicki} in order to study how the squeezing imposed by the 
expanding universe affects the entanglement between local degrees of freedom. 
We further study 
how this entanglement evolves after the inflationary era, during a radiation dominated or 
matter dominated universe.
We perform an explicit calculation of the entanglement entropy along the lines
of \cite{srednicki} for the toy model of two coupled harmonic oscillators.
We then deduce the expected features of the entropy for a non-interacting 
quantum field. Finally, we carry out an explicit computation of the entanglement
entropy for a field in $1+1$ dimensions, through which these features 
become apparent.

In the following section we summarize the relevant facts about the 
representation of a non-interacting quantum field as a collection of an 
infinite number of quantum harmonic oscillators. We distinguish between the
expansion in terms of momentum modes and the expansion in terms of
local oscillators in coordinate space. In the first case, the absence of a direct 
coupling between modes in a flat background makes the notion of entanglement more
difficult to implement. Tracing out some of them does not result in a non-trivial
reduced density matrix.
In the second case, the degrees of freedom correspond to local field
values at various points in space, which are coupled through the derivative term
in the field action. This is the starting point of the analysis in \cite{srednicki},
which we adopt. 

In section \ref{QCS} we solve for the wave function of a quantum oscillator 
whose eigenfrequency has a time dependence that results from the effective
mass term of a scalar field in an expanding background. For an inflationary 
background, we impose appropriate
initial conditions such that the wave function at early times has the standard form
for an oscillator in a flat background. This implements the assumption of a 
Bunch-Davies vacuum in this setup. 
The features of 
freezing and squeezing as a result of horizon crossing are reflected
in the properties of the wave function at late times. 

In section \ref{EntEnt} we analyse the case of two coupled oscillators in
an expanding background. We trace out one of them, determine the reduced
density matrix of the other one, and find its eigenfunctions and eigenvalues.
With this knowledge, we compute the entanglement entropy.

In section \ref{QF} we consider a large number of coupled oscillators
that result from the decomposition of a quantum field. We deduce the 
general form of the density matrix when a certain number of them are
traced out. An exact determination of the eigenfunctions of the reduced 
density matrix is not possible in this case. However, we develop 
a method that permits the numerical computation of its eigenvalues. 
Moreover, we are able to 
determine the late time dependence of the eigenvalues in an inflationary
background, which allows us to deduce the time dependence of the 
entropy. 

In section \ref{2d} we perform a numerical calculation of the entanglement entropy 
in a toy model of a field in $1+1$ dimensions, whose mass has the same time dependence
as for a field in a $(3+1)$-dimensional universe during inflation or radiation domination. 
We show how the entropy evolves from the well-known logarithmic expression for a static 
background towards a form that displays several novel features, such as 
enhancement or oscillations of the entropy.

In section \ref{concl} we present a summary of the results and our conclusions.

In the appendix \ref{appendixA} we summarize some technical points about the mode wave function
for massive fields in an expanding background and its effect on the form of the entanglement
entropy.

\section{The quantum field as a collection of quantum oscillators} \label{qoqf}

\subsection{Expanding the field in momentum modes}\label{QMS}

We consider a free scalar field in a Friedmann-Robertson-Walker (FRW) background, described by the metric 
\be 
ds^2=a^2(\tau)\left(d\tau^2 -dr^2-r^2d\Omega^2 \right)
\label{dsmetric} \ee
in terms of the conformal time $\tau$.
Through the definition $\phi(\tau,\bx)=f(\tau,\bx)/a(\tau)$,
the action can be written as
\be
S=\frac{1}{2} \int d\tau\, d^3\bx\,\left(f'^2-(\nabla f)^2
+\left( \frac{a''}{a}-a^2 m^2 \right)f^2 \right),
\label{action1} \ee
where the prime denotes differentiation with respect to the conformal time $\tau$. 
The field $f(\tau,\bx)$ has a canonically normalized kinetic term, 
i.e. its conjugate momentum is simply $f'(\tau,\bx)$.

For a de Sitter (dS) background, the scale factor is
\be
a(\tau)=-\frac{1}{H\tau},
\label{atau} \ee
with $\tau$ ranging from $-\infty$ to $0^-$. 
The action assumes the form
\be
S=\frac{1}{2} \int d\tau\, d^3\bx\,\left(f'^2-(\nabla f)^2
+\frac{2\kappa}{\tau^2} f^2 \right),
\label{action2}
\ee
where the parameter $\kappa$ is defined as
\be
\kappa=1-\frac{m^2}{2H^2}.
\label{kappa} \ee
The equation of motion is the so-called Mukhanov-Sasaki equation \cite{mukhanov,sasaki,physrep}
\be
f''-\nabla^2f-\frac{2\kappa}{\tau^2} f=0,
\label{ms} \ee
or equivalently in Fourier space
\be
f_k''+k^2f_k-\frac{2\kappa}{\tau^2}f_k=0.
\label{msf} \ee

In the quantization procedure, the solutions of eq. (\ref{msf}) can be used as mode 
functions for the expansion of the field. 
The general solution is given in terms of Bessel functions as
\be
f_k(\tau)=A_1 \, \sqrt{-\tau}\, J_\nu \left(-k\tau\right)
+A_2\, \sqrt{-\tau}\, Y_\nu \left(-k\tau\right) ,
\label{modef} \ee
where
\be
\nu = \frac{1}{2}\sqrt{1+8\kappa}
\label{nu}
\ee
and $J_n \left( x \right)$ and $Y_n \left( x \right)$ stand for the Bessel functions of the first and second kind, respectively. The Bunch-Davies vacuum corresponds to the choice
\be
A_1=-\frac{\sqrt{\pi}}{2},
\quad\quad
A_2=-\frac{\sqrt{\pi}}{2} i,
\label{bd}\ee
for which the solution at early times $(\tau \to -\infty)$
takes the standard form of a positive-frequency mode function in 
Minkowski space, up to a phase, i.e.
\be
f_k(\tau)\simeq \frac{1}{\sqrt{2k}}e^{-ik \tau}.
\label{modeappr}\ee

For a massless scalar, which corresponds to the specific value $\kappa=1$, and thus $\nu = 3/2$, the full solution reads 
\be
f_k(\tau)=\frac{1}{\sqrt{2k}}e^{-ik \tau} \left(1-\frac{i}{k\tau} \right).
\label{mode1} \ee
We have defined $A_1$, $A_2$ so that the phase vanishes at late times ($\tau \to 0^-$), when the field $\phi_k(\tau)=f_k(\tau)/a(\tau)$ becomes constant.

For a conformally coupled scalar ($\kappa=0$, or $m^2=2H^2$ and thus $\nu = 1/2$),
the full solution, up to a phase, is 
\be
f_k(\tau)=\frac{1}{\sqrt{2k}}e^{-ik \tau} 
\label{mode2} \ee
and the field $\phi_k(\tau)$ vanishes at late times.

The quantum field can be expressed as
\be
\hat{f}(\tau,\bx)=\int \frac{d^3\bk}{(2\pi)^{3/2}}
\left[f_k(\tau)\hat{a}_\bk+f^*_k(\tau)\hat{a}^\dagger_\bk\right] e^{i\bk\cdot\bx}
\label{qff} \ee
where $\hat{a}^\dagger_\bk$, $\hat{a}_\bk$ are standard 
creation and annihilation operators, i.e. they obey the commutation relations $\left[ \hat{a}_\bk , \hat{a}^\dagger_{\bk '} \right] = \delta^3 \left( \bk + \bk ' \right)$.
The expectation value of the field vanishes, while its variance can be
calculated to be
\be
\langle \hat{f}^2 \rangle=\int d\ln k\, \frac{k^3}{2\pi^2}|f_k(\tau)|^2,
\label{varf} \ee
with
\be
|f_k(\tau)|^2=-\frac{\pi}{4} \tau
\left[
J_\nu^2 \left(-k\tau\right) +
Y_\nu^2 \left(-k\tau\right)
\right].
\label{fsq} \ee
The variance of the conjugate field $\hat{\pi}(\tau,\bx)=\partial\hat{f}(\tau,\bx)/\partial \tau$
is
\be
\langle \hat{\pi}^2 \rangle=\int d\ln k\, \frac{k^3}{2\pi^2}|f'_k(\tau)|^2.
\label{varp} \ee

For a massless field we have $\kappa=1$ and 
$f_k(\tau)$ is given by eq. (\ref{mode1}).
For superhorizon modes with $k\tau\to 0^-$ the second term in eq. (\ref{mode1}) dominates. 
If only this term is retained one obtains
\be
f_k(\tau)=-\frac{i}{\sqrt{2}k^{3/2}}\frac{1}{\tau}
=-\tau  f'_k(\tau)
\label{fkasy} \ee
and, therefore, 
\be
\hat{\pi}(\tau,\bx)=-\frac{1}{\tau}\hat{f}(\tau,\bx).
\label{fhasy} \ee
The fact that the dominant term of the field and the dominant term of its conjugate momentum commute indicates that for most of
its properties it can be 
viewed as a classical stochastic field instead of a quantum one
\cite{albrecht}.
The $k$-dependence of the mode function results in the standard
scale-invariant power spectrum. 

However, there are quantities of purely quantum origin, for which a classical
description is inadequate. A characteristic example is quantum entanglement, which is the focal point of this work. 
It is important to realize that the full quantum field and its conjugate always 
obey the canonical commutation
relation. This is guaranteed by the presence of the subleading first term 
in the mode function (\ref{mode1}). As a result, the entanglement entropy does not necessarily
vanish for superhorizon modes. On the contrary,
eq. (\ref{fkasy}) implies that each mode evolves into 
a squeezed state, for which the quantum effects get enhanced \cite{Grishchuk}. 
This becomes apparent if one observes that, 
for a field composed only of superhorizon modes,
the ratio of the variance of $\hat{f}$ to the variance of its conjugate
$\hat{\pi}$ vanishes. We shall discuss this feature in detail in the following
sections.

\subsection{Expanding the field in coordinate space}
\label{subsec:exp_coordinate}

Based on the formalism of the previous section, one may consider the 
momentum-space entanglement 
between high and low-momentum modes,
such as between modes with physical momenta below and above
the Hubble scale $H$ \cite{momentumspace1,momentumspace2}.
For a free field described by a quadratic action, where the momentum modes do not interact, the entanglement entropy
would vanish, as long as the initial state can be written as a tensor product of one state for each momentum mode, as in the Minkowski vacuum.
Since each mode evolves independently, the reduced 
density matrix, when some modes are traced over,
would be one of a pure state, namely the state of the modes which have not been traced out. 

We are interested instead in the entanglement 
between degrees of freedom
localized within two spatial regions separated by an entangling surface. This point of view is also more interesting in the context of understanding 
the scaling properties of entanglement entropy, 
namely the dominance of an area law term, which indicates a 
similarity with black hole entropy.
For a dS background
one may consider the entanglement between the interior of a horizon-size region
of radius $1/H$ and the exterior. 
In such an approach one has to trace over the degrees of freedom in the 
interior of the entangling surface. The reduced density matrix would not 
correspond to a pure state and the entanglement entropy would be non-zero.
This is the classic approach by Srednicki for flat space in \cite{srednicki}, which we would like to 
consider for the case of a time-dependent background.

For spherical entangling surfaces, it is convenient to define the spherical moments of the field and its momentum as
\begin{equation}
{f _{lm}} \left( r \right) = r \int {d\Omega \, {Y_{lm}}\left( {\theta ,\varphi } \right) f \left( {\bx} \right)} , \quad\quad
{\pi _{lm}}\left( r \right) = r \int {d\Omega \, {Y_{lm}}\left( {\theta ,\varphi } \right) \pi \left( 
{\bx} \right)} ,\label{eq:moments}
\end{equation}
where ${Y_{lm}}$ are real spherical harmonics. The radial coordinate can be discretized by
introducing a lattice of concentric spherical shells
with radii $r_j = j \epsilon$, where $j$ is an integer obeying $1 \le j \le N$. 
The radial distance between successive shells introduces an ultraviolet (UV) 
cutoff equal to $1/\epsilon$, while 
the total size of the lattice $L=N \epsilon$ sets an infrared (IR) cutoff equal to $1/L$. 
By defining the discretized degrees of freedom as 
\begin{equation}
{f _{lm}}\left( {j\epsilon} \right) \to {f _{lm,j}} , \quad\quad
{\pi _{lm}}\left( {j\epsilon} \right) \to \frac{{{\pi _{lm,j}}}}{\epsilon}, 
\end{equation}
so they are canonically commuting,
we arrive at the Hamiltonian 
\be
H = \frac{1}{{2\epsilon}}\sum\limits_{l,m} {\sum\limits_{j = 1}^N {\left[ {\pi _{lm,j}^2 + {{\left( {j + \frac{1}{2}} \right)}^2}{{\left( {\frac{{{f_{lm,j + 1}}}}{{j + 1}} - \frac{{{f_{lm,j}}}}{j}} \right)}^2} + \left( \frac{{l\left( {l + 1} \right)}}{{{j^2}}}-\frac{2\kappa}{(\tau/\epsilon)^2} \right)f_{lm,j}^2} \right]} } ,
\label{eq:Hamiltonian_discretized}
\ee
where we have assumed a dS background. 
Throughout the paper all dimensionful quantities are assumed to be given in units
of the comoving lattice spacing $\epsilon$, with $1/\epsilon$ acting as a UV cutoff.
This normalization corresponds to setting $\epsilon=1$. We denote $\epsilon$ explicitly
whenever the dependence on the UV cutoff is crucial. 

The problem of the entanglement entropy between the interior and exterior of a sphere of
radius $R$ can now be addressed by computing the density matrix for the above discretized harmonic system, via
tracing out the oscillators with $j \epsilon <R $ in order to compute the 
reduced matrix for the degrees of freedom outside the entangling surface (or vice versa), from 
which the entanglement entropy can be calculated. 
For the vacuum of the theory, the state of the
system of oscillators must be assumed to be the product of the `ground states' 
of the modes that
diagonalize the Hamiltonian. 
The assumption of a Bunch-Davies vacuum implies that 
as a `ground state' of a mode we must define the solution of 
the time-dependent Schr\"{o}dinger equation for this mode, 
which reduces to the usual simple harmonic oscillator ground state 
as $\tau \to - \infty$. 

Thus, one must determine first the eigenmodes of this system of 
coupled oscillators. The wave function of each mode depends on a linear combination of the 
various $f_{lm,j}$, i.e. the corresponding canonical coordinate. Since modes with 
different $l$ and $m$ indices do not mix, each eigenfunction actually involves one set of 
$(l,m)$. The eigenvalues of each mode correspond to the eigenfrequencies of the system, which
vary from high values $\sim 1/\epsilon$ to low values $\sim 1/L$. The effect of the expanding background is encoded in the term $-{2\kappa}/{(\tau/\epsilon)^2} f_{lm,j}^2$. This term is identical for all $(l,m)$. As such, it does not mix different $(l,m)$ 
and it does not affect the rotation matrix which acts on the vector of the coordinates and diagonalizes the Hamiltonian. 
In other words, the canonical coordinates are identical to those in 
the absence of this term, i.e. for the Minkowski vacuum. 
Furthermore, the aforementioned rotation leaves this term invariant. In other words, if the discretized Hamiltonian of the free field in Minkoswki space can be written as
\be
H = \frac{1}{{2\epsilon}}\sum\limits_{l,m} {\sum\limits_{j = 1}^N {\left[ {\tilde{\pi} _{lm,j}^2 + \left( \omega_{lm,j}^2+m^2 \right) 
 \tilde{f}_{lm,j}^2} \right]} } ,
\ee
 where $\tilde{f}_{lm,j}$ 
are the canonical coordinates, then the discretized Hamiltonian for the free field in 
an inflationary background will be
\be
H = \frac{1}{{2\epsilon}}\sum\limits_{l,m} {\sum\limits_{j = 1}^N {\left[ {\tilde{\pi} _{lm,j}^2 + \left( \omega_{lm,j}^2-\frac{2\kappa}{(\tau/\epsilon)^2} \right)\tilde{f}_{lm,j}^2} \right]} } .
\ee
It follows that we need to solve for the harmonic oscillator with 
a time-dependent eigenfrequency of the form $\omega_0-2\kappa/\tau^2$. in order to find the `ground states' of the field modes.

It was shown by Srednicki \cite{srednicki} that,
in a static background, the entanglement entropy scales with the 
area of the entangling surface. This result also sheds light on the influence of the 
UV sector of the system and the role of the UV cutoff $1/\epsilon$.
The absence of an intrinsic energy scale other that $1/\epsilon$ for a massless theory implies
that the entanglement entropy must depend on the ratio $R/\epsilon$. This is also clear from
the form of the Hamiltonian (\ref{eq:Hamiltonian_discretized}).  
The divergence of the entropy for $\epsilon \to 0$ indicates that the UV sector gives the
dominant contribution. The usual interpretation is that the entropy results mainly from the 
strong entanglement of the short distance modes on either side of
the entangling surface, while there is no significant volume contribution.  

The above constraint does not apply 
to an expanding background, as there is a new energy scale set by the expansion rate
$H$. In the dS case, UV modes can be stretched by the expansion beyond the horizon, and their
form may be altered drastically, as we discussed in the previous section. The wave functions
describing the quantum properties of such modes will not be the typical simple harmonic oscillator ones. 
In the following sections we examine in detail the form of the oscillator wave function in 
various expanding backgrounds and the implications for the entanglement entropy.

\section{The oscillator wave function in an expanding background}\label{QCS}

\subsection{De Sitter background}

A basic quantity of interest is the wave function of an oscillator with an effective eigenfrequency
composed of a constant
contribution originating in the spatial derivative term in eq. (\ref{action2})
and a time-dependent contribution arising through the expanding background. 
More specifically, an effective time-dependent eigenfrequency of the form
\be
\omega^2(\tau)=\omega^2_0-\frac{2\kappa}{\tau^2}
\label{tdmass} \ee
corresponds to an oscillator in a dS background. (Compare with eqs. (\ref{action2}), (\ref{kappa}).) The conformal time $\tau$ ranges from $-\infty$ to $0^-$. 
The solution of the Schr\"odinger equation can be achieved in several steps \cite{guerrero}, following the Lewis–Riesenfeld
method \cite{lewis1,lewis2}.
First, one must find the general solution of the Ermakov equation 
\be
b''(\tau)+\omega^2(\tau)b(\tau)=\frac{\omega^2_0}{b^3(\tau)}.
\label{ermakoveq} \ee 
This can be obtained from two linearly independent solutions
$y_1(\tau)$, $y_2(\tau)$ of the 
linear equation
\be
y''(\tau)+\omega^2(\tau)y(\tau)=0,
\label{yeq} \ee 
as 
\be
b^2(\tau)=c_1\, y_1^2(\tau)+c_2\, y_2^2(\tau)+2 c_3\, y_1(\tau) y_2(\tau),
\label{ermakovsol} \ee
where the constants $c_1$, $c_2$ and $c_3$ must obey
\be
c_1 c_2-c_3^2=A,
\label{cconstr} \ee
with $A$ a constant that depends on the form of $\omega(\tau)$.
For the problem at hand, the two independent solutions of eq. (\ref{yeq}) are 
$y_1 = \sqrt{-\tau}\, J_\nu \left(-\omega_0\tau \right)$
 and 
 $y_2 = \sqrt{-\tau}\, Y_\nu \left(-\omega_0\tau \right)$, where $\nu$ is given by 
 eq. \eqref{nu}
 and $A=\pi^2\omega_0^2/4$.
 The values of $c_1$, $c_2$ are fixed by imposing appropriate initial conditions.
 
\begin{figure}[t!]
\centering
\includegraphics[width=0.6\textwidth]{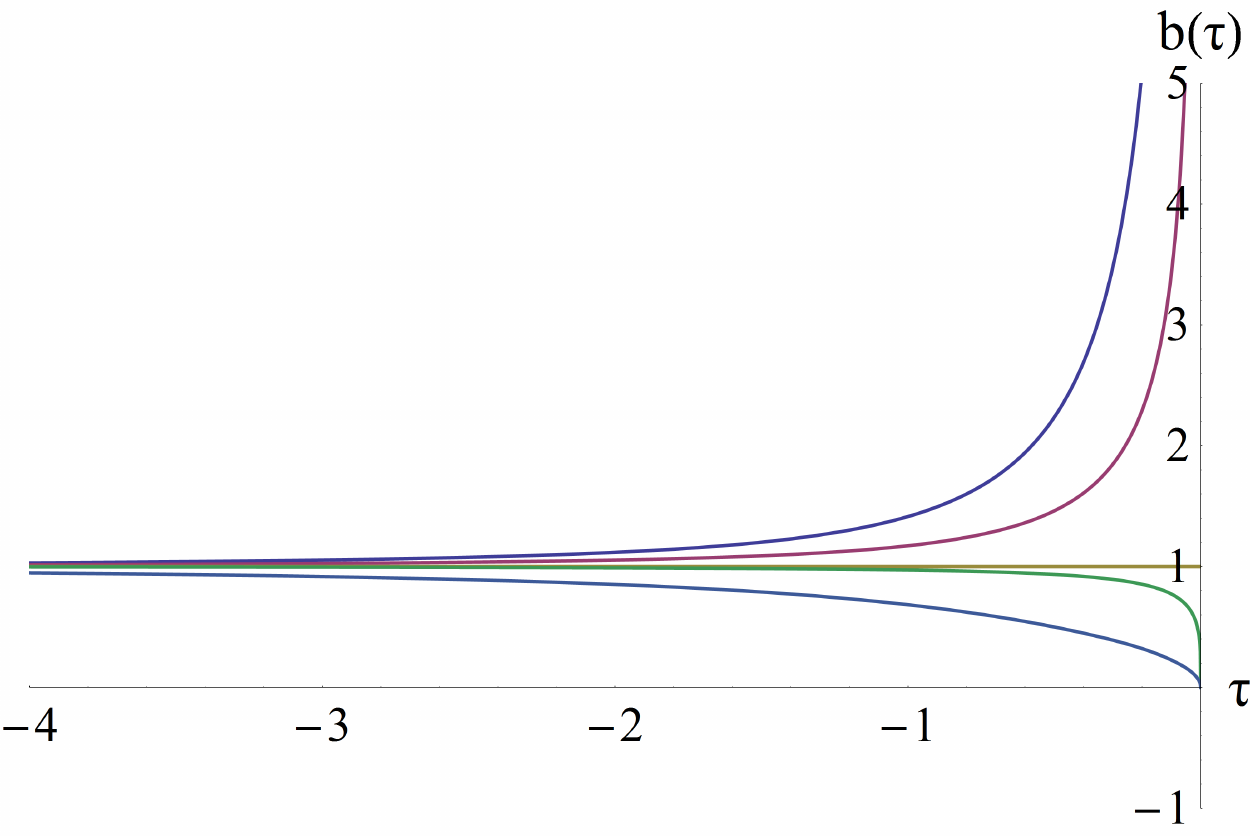} 
\caption{
The form of the function $b(\tau)$ for $\omega_0=1$ and $\kappa=$1, 0.5, 0, $-0.1$, $-2$ (from top to bottom). 
}
\label{fig1}
\end{figure}

 The assumption of a Bunch-Davies vacuum implies that we  
need the solutions of the Schr\"odinger equation that reduce to the
standard solutions of an oscillator of constant frequency $\omega_0$ for $\tau\to-\infty$. 
We require that the form of $b(\tau)$ become trivial for $\tau\to -\infty$,
so that we are led to a solution that corresponds to the standard harmonic oscillator.
This imposes the conditions
$b(\tau)\to 1$, $b'(\tau)\to 0$ for $\tau\to -\infty$, 
which are satisfied for $c_1=c_2=\pi\omega_0/2$ and, therefore, $c_3=0$.
In this way we obtain
\be
b^2(\tau)=-\frac{\pi}{2}\omega_0 \tau
\left(
J^2_\nu \left(-\omega_0\tau \right) +
Y^2_\nu \left(-\omega_0\tau \right)
\right).
\label{bsolfin} \ee
The Bessel functions can be continued to imaginary order for $\kappa<-1/8$, with
$b(\tau)$ remaining real.
In figure \ref{fig1} we depict the form of $b(\tau)$ for various values of $\kappa$.
For $\kappa>-1/8$ and $\tau \to 0^-$ we have
\be
b(\tau) \simeq \frac{\Gamma\left( \nu \right)}{\sqrt{\pi}}
\left(\frac{-\omega_0 \tau}{2} \right)^{\frac{1}{2} - \nu} .
\label{appr1} \ee
The asymptotic form of $b(\tau)$ plays a crucial role in the growth of the entanglement
entropy at late times \cite{Chandran}.
For $\kappa=0$ we have $b(\tau)=1$ at all times. For $\kappa<-1/8$ and $\tau \to 0^-$ 
we have\footnote{The apparent discontinuity at $\kappa=-1/8$ of formulae \eqref{appr1} and \eqref{appr2} by a factor of 2 is due to the fact that formula \eqref{appr1} ceases being a good approximation for $\kappa \to -1/8$,
as the contribution to the function $b(\tau)$ from the Bessel function of the first kind becomes equally important to that of the Bessel function of the second kind. In this regime, formula \eqref{appr1} must be corrected to
\begin{equation}
b(\tau) \simeq \left(\frac{-\omega_0 \tau}{2 \pi} \right)^{\frac{1}{2}} \left( \Gamma^2\left( \nu \right)
\left(\frac{-\omega_0 \tau}{2} \right)^{- 2 \nu} + 2 \Gamma\left( \nu \right) \Gamma\left( - \nu \right)
\cos \left( \pi \nu \right) + \Gamma^2\left( - \nu \right)
\left(\frac{-\omega_0 \tau}{2} \right)^{2 \nu} \right)^{\frac{1}{2}}.
\end{equation}}
\begin{equation}
b(\tau) \simeq \frac{\left|\Gamma\left(i | \nu | \right)\right|}{\sqrt{\pi}}
\left(-\omega_0 \tau \right)^{\frac{1}{2}} \left( \cosh \left( \pi | \nu | \right) + \cos \left( 2 | \nu | \ln \frac{-\omega_0 \tau}{2} - 2 {\rm arg} \Gamma\left(i | \nu | \right) \right) \right)^{\frac{1}{2}} .
\label{appr2} 
\end{equation}
More details about the form of the function $b(\tau)$ are given in appendix \ref{appendixA}.

The solution of the Schr\"odinger equation can now be expressed as
\be
F(\tau,f)=\frac{1}{\sqrt{b(\tau)}}\,\exp\left({\frac{i}{2}\frac{b'(\tau)}{b(\tau)}f^2}\right)\,
F^0\left(\int \frac{d\tau}{b^2(\tau)},\frac{f}{b(\tau)}\right),
\label{solschrod} \ee
where $F^0(\tau,f)$ is a solution of the standard simple harmonic oscillator with
constant frequency $\omega_0$, namely a linear combination of the wave functions
\be
F^0_{n}(\tau,f)=\frac{1}{\sqrt{2^n n!}}\left(\frac{\omega_0}{\pi} \right)^{1/4} 
\exp\left(-\frac{1}{2}\omega_0 f^2 \right) 
H_n\left(\sqrt{\omega_0}f \right)
\exp \left(-i\left(n+\frac{1}{2} \right)\omega_0\tau \right),
\label{quantharm} \ee
with $H_n(x)$, $n=0,1,2,...$ the Hermite polynomials.
The integral of $1/b^2(t)$ cannot be obtained in closed analytic form for 
general $\kappa$. For this reason we consider values of $\kappa$ of particular interest:
\begin{itemize}
\item
For $\kappa=0$ (conformally coupled scalar with $m^2=2H^2$) we have $F(\tau,f)=F^0(\tau,f)$, so
that the oscillator evolves  
as a standard quantum oscillator in Minkowski space.
\item
For a massless scalar with $m=0$ and $\kappa=1$, we find
\be
b(\tau)=\sqrt{1+\frac{1}{\omega_0^2\tau^2}},
~~~~~~~~~~~~
\int \frac{d\tau}{b^2(\tau)}=\tau+\frac{1}{\omega_0}
\left(\tan^{-1}(-\omega_0\tau)-\frac{\pi}{2} \right),
\label{b1} \ee
where we have chosen limits such that the integrated function is approximately equal to $\tau$
for $\tau\to -\infty$. The wave function becomes a linear combination of the wave functions
\begin{multline}
F_n(\tau,f)=\frac{1}{\sqrt{2^n n!}}\left(\frac{\omega_0}{\pi}\frac{1}{1+\frac{1}{\omega_0^2\tau^2}} \right)^{1/4} 
\exp\left(-\frac{1}{2}\left( \omega_0-\frac{1}{i \tau+\omega_0\tau^2} \right) f^2 \right) \\
\times H_n\left(\sqrt{\frac{\omega_0}{1+\frac{1}{\omega_0^2\tau^2}}}f \right) \exp \left(-i\left( n+\frac{1}{2} \right)
\left( \omega_0\tau -\tan^{-1}(\omega_0\tau)-\frac{\pi}{2}\right) \right).
\label{wavefb2}  \end{multline}
These are normalized solutions of the Schr\"odinger equation that
reduce to the standard wave functions $F^0_{n}(\tau,f)$ of eq. (\ref{quantharm}) for $\tau\to-\infty$.
\end{itemize}

\begin{figure}[t!]
\centering
\includegraphics[width=0.48\textwidth]{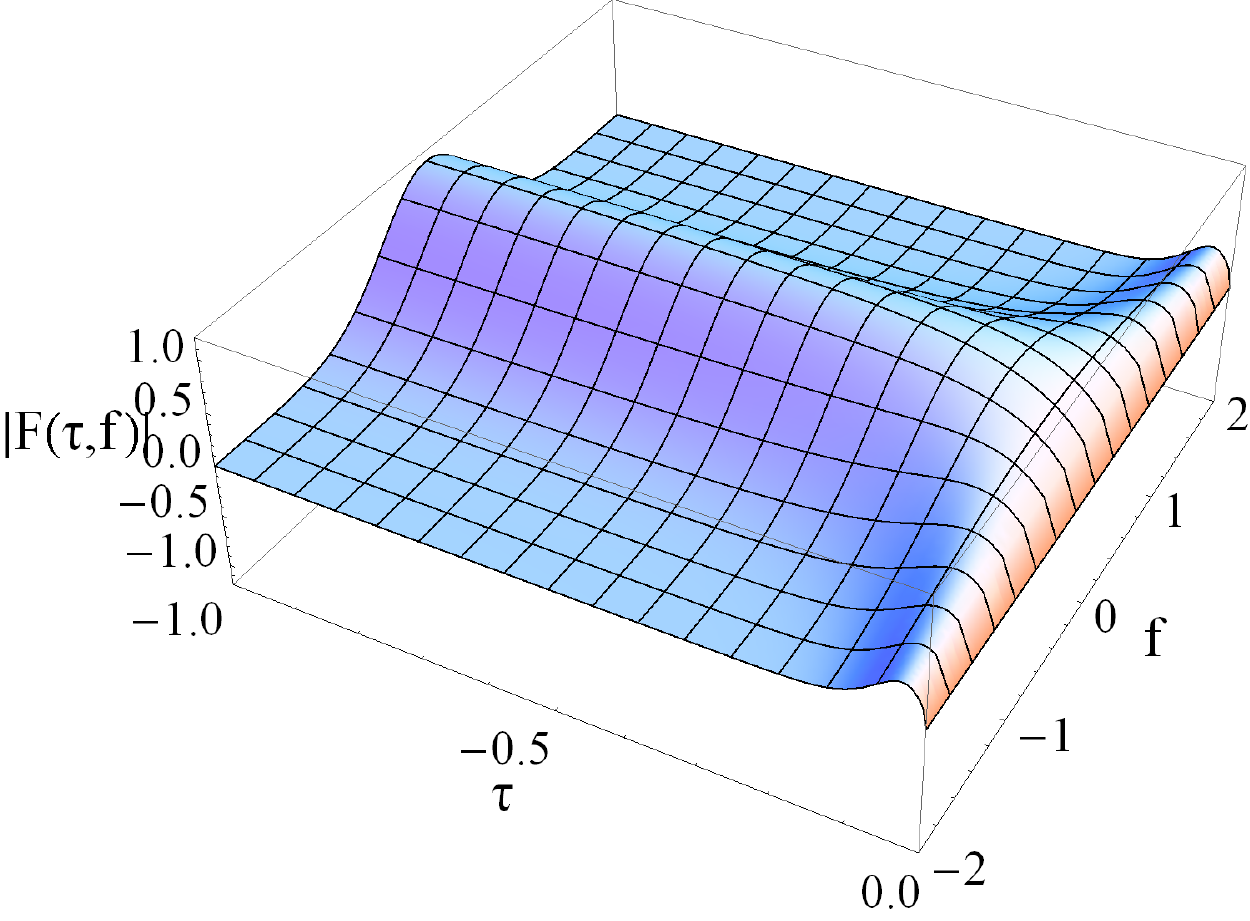} 
\includegraphics[width=0.48\textwidth]{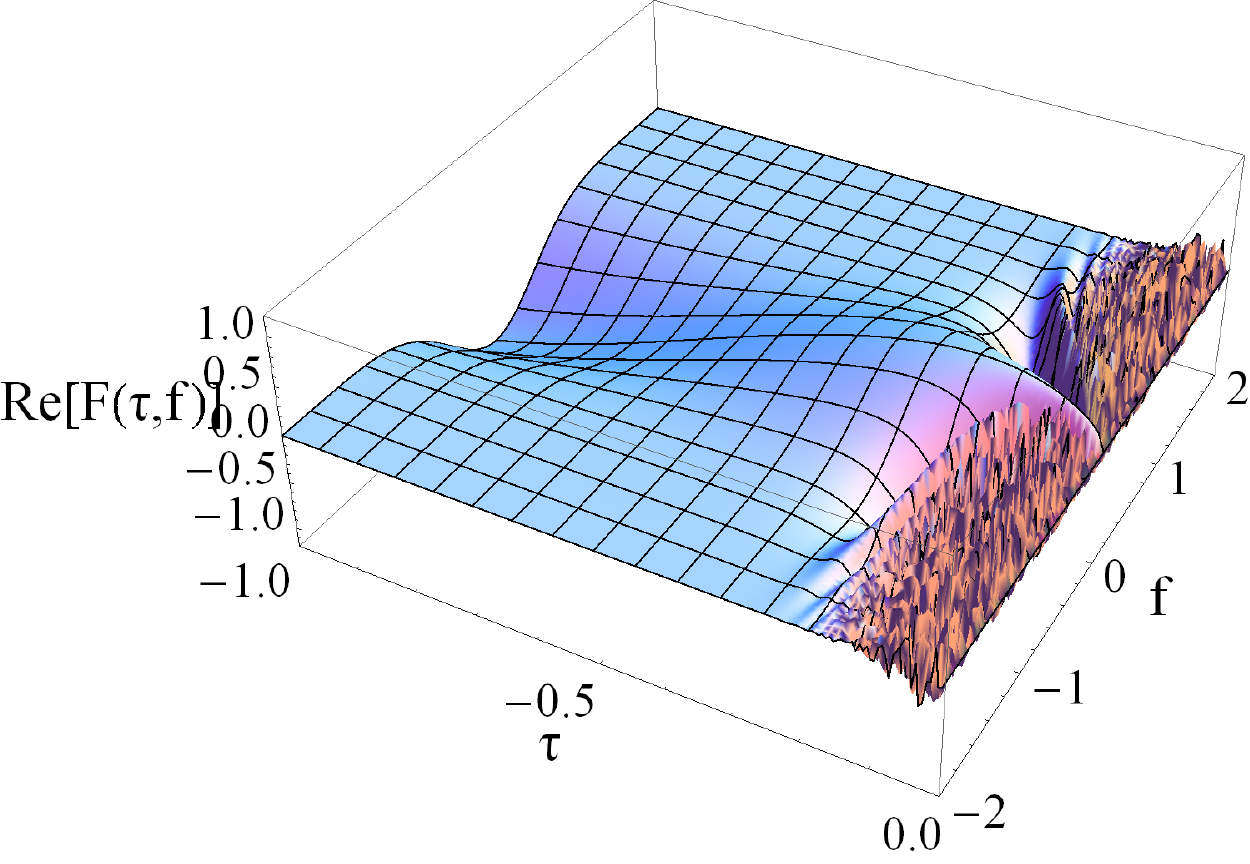} 
\caption{
The amplitude and the real part of the `ground-state' wave function, given by eq. (\ref{wavefb2}) with
$n=0$ and $\omega_0=5$.
}
\label{fig2}
\end{figure}

The amplitude of the `ground-state' wave function with $n=0$ 
and $\omega_0=1$ is depicted in the left plot of figure \ref{fig2}. 
It is apparent that the expansion of the background results in the spreading of the
wave function.  The amplitude has the form of a Gaussian with width $\sqrt{\omega}/b(\tau)$.
The maximum drops, so that the normalization remains constant.
There is a visible transition at a critical time $\tau_c \simeq -1/\omega_0$ from the 
solution with constant amplitude, as in the static case, to a configuration with a time-dependent
amplitude. By making use of eq. (\ref{atau}) we can define the time $\tau_c$ through the relation 
\be
\frac{\omega_0}{a(\tau_c)}=H.
\label{exit} \ee
The oscillatory $f$-independent pattern of the wave function with respect to time, induced by the 
last factor in eq. (\ref{wavefb2}), ceases for $\tau\to 0^-$.
This is the analogue of the process of horizon exit in this language, with the 
subsequent freezing of fluctuations.  
Despite the rather simple evolution of the amplitude, 
the full wave function is much more complicated, as is apparent 
from the right plot of figure \ref{fig2}, in which its real part is depicted.
A strong time dependence is induced by the imaginary exponent in the second factor of eq. (\ref{wavefb2}), 
arising from the exponent $i (b'(\tau)/b(\tau))f^2/2$ in eq. (\ref{solschrod}), 
which diverges for $\tau\to 0^-$. As a result, the quantum properties of this state become enhanced.

For the wave function (\ref{solschrod}) corresponding to a `ground state',
it is instructive to compute the variance of the conjugate 
operators
$\hat{f}$ and $\hat{\pi}=-i \partial/\partial f$. 
We have $\langle \hat{f} \rangle=\langle \hat{\pi} \rangle=0$ and 
\be
 \langle \hat{f}^2 \rangle\equiv \left( \Delta f\right)^2= \frac{b^2(\tau)}{2\omega_0},
\qquad
 \langle \hat{\pi}^2 \rangle\equiv \left( \Delta\pi \right)^2= \frac{\omega_0}{2b^2(\tau)}+ \frac{b'^2(\tau)}{2\omega_0}.
\label{varfvarp} \ee
For $\kappa>0$ and $\tau\to -\infty$, 
we have $\Delta f\to 1/\sqrt{2\omega_0}$ and $\Delta \pi\to \sqrt{\omega_0/2}$, so
that $\Delta f\, \Delta \pi\to 1/2$. On the other hand, when $b(\tau)$ diverges 
for $\tau\to 0^-$, we have
 $\Delta f=b(\tau)/\sqrt{2\omega_0}\to \infty$ and 
 $\Delta \pi  \simeq |b'(\tau)|/\sqrt{2\omega_0}\to \infty$.
The product $\Delta f\, \Delta \pi$ diverges in this limit, indicating that the 
quantum nature of the state becomes more prominent.
More specifically, for the case of a dS background, eq. (\ref{bsolfin}) implies that
\be
\left( \Delta f\right)^2= \frac{b^2(\tau)}{2\omega_0}
=-\frac{\pi}{4}\tau
\left(
J^2_\nu \left(-\omega_0\tau \right) +
Y^2_\nu \left(-\omega_0\tau \right)
\right).
\label{df22} \ee
The analogy with eqs. (\ref{varf}), (\ref{fsq}) is apparent.
For $\kappa>0$ and  $\tau\to 0^-$, we have $\Delta f/ \Delta \pi\simeq b(\tau)/|b'(\tau)|\sim |\tau| \to 0$. Even though both $\Delta f$ and $\Delta \pi$ diverge, the uncertainty 
is much larger in the determination of the momentum. 

Gaussian states that are not minimal uncertainty states are not new in the 
literature. In the context of the standard quantum oscillator
they correspond to the so-called squeezed states. 
These are states which are characterized by a Gaussian amplitude and a phase 
that is a quadratic function of the coordinate. 
The squeezed states are not minimal uncertainty states, while
the uncertainties of position and momentum are generally unbalanced. 
This is exactly the form of our wave functions, which
reduce to coherent states when the quadratic phase vanishes, resulting in minimal 
uncertainty states. 
The deviation from a coherent state is typically parametrized via a so-called squeezing 
parameter. In our case, the wave function emerges as the ground state of the 
time-independent oscillator, i.e. a special coherent and thus a minimal uncertainty state,
and evolves towards an increasingly squeezed state as $\tau \to 0^-$. 
Since uncertainties increase, several quantum-mechanical characteristics 
of the state are enhanced.

\subsection{Radiation and matter domination}

Even though the discussion of the quantum oscillator in a general FRW background is 
complicated, the most interesting cases are amenable to an analytical treatment. 
In a radiation-dominated Universe the scale factor evolves $\sim \tau$ for large
positive $\tau$, while during matter domination it evolves $\sim \tau^2$. 
In the standard cosmological scenario, early-time inflation is followed by 
a period of radiation domination (RD), which is subsequently followed by a period
of matter domination (MD). It is particularly interesting to understand how a 
squeezed state formed during the inflationary epoch would evolve during these 
subsequent stages. In order to keep the analytical expressions as simple as 
possible, we shall consider two idealized scenarios. In the first one, 
there is an instantaneous transition from inflation to radiation domination, while
in the second, inflation is followed by matter domination. The 
scenario inflation-RD-MD can be analysed along the same lines, but it results in 
much more complicated analytical expressions and the same qualitative behaviour.

Assuming that the transition from a dS to either a RD or MD background 
takes place at $\tau=\tau_0$, we define the scale factors 
\begin{equation}
a(\tau) = \begin{cases}
[1-H(\tau-\tau_0)]^{-1}, &\text{dS},\\
1+H(\tau-\tau_0), &\text{RD},\\
\left[ 1 +\frac{1}{2} H (\tau - \tau_0) \right]^2, &\text{MD},
\end{cases}
\label{scalefactors}
\end{equation}
so that
\begin{equation}
    a_\text{dS}(\tau_0)=a_\text{RD}(\tau_0)=a_\text{MD}(\tau_0)=1.
\end{equation}
The Hubble parameter is continuous through the time $\tau_0$. However, the 
contribution $a''/a$ to the effective frequency is discontinuous.
We limit our discussion to oscillators arising in the case of a massless field, 
for which the effective frequency is of the form
\be
\omega^2(s)=\omega^2_0-\frac{2\kappa}{s^2},
\label{tdfreq} \ee
with
$\kappa=1$ for dS and MD and $\kappa=0$ for RD, and
\begin{equation}
\label{ses}
s(\tau) = \begin{cases}
\tau-\tau_0-\frac{1}{H}, &\tau<\tau_0,\quad \text{dS},\\
\tau-\tau_0+\frac{2}{H}, &\tau >\tau_0,\quad \text{MD}.
\end{cases}
\end{equation}
The wave function of the quantum oscillator can be computed along the lines of
the previous section, demanding continuity at $\tau = \tau_0$. 

Since during radiation domination we have $\kappa = 0$, 
the wave functions of the various modes evolve as wave functions of the standard harmonic 
oscillator. 
The same holds for the late-time evolution in the matter domination era, 
since the effective eigenfrequency of each mode is 
$\omega_0 - 2 \kappa/\tau^2 \simeq \omega_0$. 
We know that, during the inflationary era, the wave function of each field mode 
evolves from the 
ground state to a squeezed state of the simple harmonic oscillator. 
Therefore, the wave function of each mode during the subsequent RD era, 
or at late times in the MD era, also evolves like a squeezed state of the simple harmonic oscillator. 
It is well known that such squeezed states 
evolve periodically, with a period set by the eigenfrequency. 
Twice in this period the squeezed states become minimal uncertainty states, 
with unbalanced uncertainties for position and momentum. 
In general, the product of the uncertainties oscillates between $1/2$  
and a maximal value. 
As a result, we expect that the evolution of each field mode
during the RD era, 
or at late times of the MD era, will be periodic, and that   
the contribution to entanglement by each mode will be fluctuating but periodic. 
The eigenfrequencies of the various modes do not have rational ratios.
Therefore, we should not expect a periodic fluctuation of the total 
entanglement entropy, unless 
the dynamics of the overall harmonic system have been specifically selected. 
However, we expect that the entanglement entropy 
fluctuates between a minimal and maximum value and it does not have an overall increasing 
or decreasing trend.

For a dS background, the form of $b(\tau)$ is given by the first of eqs. (\ref{b1})
with the replacement of $\tau$ by $s(\tau)$. This gives
\be
b^2_{\rm dS}(\tau)=1+\frac{1}{\omega_0^2\left( \tau-\tau_0-\frac{1}{H} \right)^2},
\quad
\tau<\tau_0.
\label{bdS2sf} \ee

In the radiation dominated era the effective frequency of the quantum oscillator is
time-independent. However, the oscillator cannot be in an energy eigenstate, as this
is not consistent with the continuity of the wave function at the point of transition
from the previous inflationary era. A solution can be obtained if we assume that the 
wave function retains the 
form of eq. (\ref{solschrod}) and we determine the function 
$b(\tau)$ by satisfying eqs. (\ref{ermakoveq})-(\ref{cconstr}). 
The two independent solutions of eq. (\ref{yeq}) are now 
$\cos(\omega_0 \tau)$ and $\sin(\omega_0 \tau)$, while the constants 
$c_1$, $c_2$, $c_3$ satisfy $c_1 c_2-c_3^2=1$. We next require the continuity of
$b(\tau)$ and its derivative at the time $\tau_0$ of transition from the 
dS to the RD background, which fixes the values of $c_1$, $c_2$. In this way we obtain
\be
b^2_{\rm RD}(\tau)=c_1\cos^2(\omega_0\tau)+c_2\sin^2(\omega_0\tau)
+ 2 c_3 \cos(\omega_0\tau)\sin(\omega_0\tau) ,
\label{brd} 
\ee
where
\be
\begin{split}
c_1&=\frac{H^4+2\omega_0^4+H^2(2\omega_0^2-H^2)\cos(2\omega_0 \tau_0)-2H^3\omega_0\sin(2\omega_0 \tau_0)}{2\omega_0^4} , \\
c_2&=\frac{H^4+2\omega_0^4-H^2(2\omega_0^2-H^2)\cos(2\omega_0 \tau_0)+2H^3\omega_0\sin(2\omega_0 \tau_0)}{2\omega_0^4} , \\
c_3&=\frac{2H^3 \omega_0\cos(2\omega_0 \tau_0)+ H^2(2\omega_0^2-H^2)
\sin(2\omega_0 \tau_0)}{2\omega_0^4}.
\end{split}
\ee
The above expressions can be simplified to 
\be
b^2_{\rm RD}(\tau)=
1+\frac{H^4}{2\omega_0^4}
+\left(\frac{H^2}{\omega_0^2}-\frac{H^4}{2\omega_0^4} \right)
\cos(2\omega_0 (\tau-\tau_0))
+ \frac{H^3}{\omega_0^3}\sin(2\omega_0 (\tau-\tau_0)),
\quad
\tau>\tau_0.
\label{bRD2f} \ee

If inflation is followed by matter domination, the 
two independent solutions of eq. (\ref{yeq}) can be taken as
$ \sqrt{s(\tau)}\, J_{\frac{3}{2}}(\omega_0 s(\tau))$
 and 
 $ \sqrt{s(\tau)}\, Y_{\frac{3}{2}}(\omega_0s(\tau))$.
 The constants $c_1$, $c_2$, $c_3$ satisfy eq. (\ref{cconstr})
with $A=\pi^2\omega_0^2/4$. 
The values of $c_1$, $c_2$ are now fixed by the continuity 
of $b(\tau)$ at $\tau_0$. 
In this way we obtain
\be
b^2_{\rm MD}(\tau)=
c_1 s(\tau)\, J_{\frac{3}{2}}^2(\omega_0s(\tau))
+c_2 s(\tau) \, Y_{\frac{3}{2}}^2(\omega_0s(\tau))
+ 2 c_3 s(\tau)\, J_{\frac{3}{2}}(\omega_0s(\tau))\, Y_{\frac{3}{2}}(\omega_0s(\tau)) ,\\
~~
\label{bmd}
\ee
where
\be
\begin{split}
c_1&= \pi 
\frac{
9H^6+9H^4\omega_0^2+32\omega_0^6
}{64\omega_0^5}
\\
&+
3\pi H^2 
\frac{
(3H^4- 21\omega_0^2 H^2+8 \omega_0^4)\cos(4\omega_0/H)
+4H\omega_0(3H^2-5\omega^2_0)\sin(4\omega_0/H)
}{64\omega_0^5}
\\
c_2&= \pi 
\frac{
9H^6+9H^4\omega_0^2+32\omega_0^6
}{64\omega_0^5}
\\
&-
3\pi H^2 
\frac{
(3H^4- 21\omega_0^2 H^2+8 \omega_0^4)\cos(4\omega_0/H)
+4H\omega_0(3H^2-5\omega^2_0)\sin(4\omega_0/H)
}{64\omega_0^5}
\\
c_3&=3\pi H^2 
\frac{
-4H \omega_0(3H^2- 5\omega_0^2)\cos(4\omega_0/H)
+(3H^4-21 H^2 \omega^2_0+8\omega_0^4)\sin(4\omega_0/H)
}{64\omega_0^5}.
\end{split}
\ee
Despite the apparent complexity of the above expressions, $b_{\rm MD}(\tau)$ can be 
simplified considerably. For large $\tau$ it takes a form similar to eq. (\ref{bRD2f}) and
reads 
\begin{multline}
b^2_{\rm MD}(\tau)=
1+\frac{9H^4}{32\omega_0^4}+\frac{9H^6}{32\omega_0^6}
+\left(\frac{3H^2}{4\omega_0^2}-\frac{63H^4}{32\omega_0^4}+\frac{9H^6}{32\omega_0^6} \right)
\cos(2\omega_0 (\tau-\tau_0)) \\
+\left( \frac{15H^3}{8\omega_0^3}- \frac{9H^5}{8\omega_0^5} \right) \sin(2\omega_0 (\tau-\tau_0)), \quad
\tau>\tau_0.
\label{bMD2fas} \end{multline}

\begin{figure}[t!]
\centering
\includegraphics[width=0.45\textwidth]{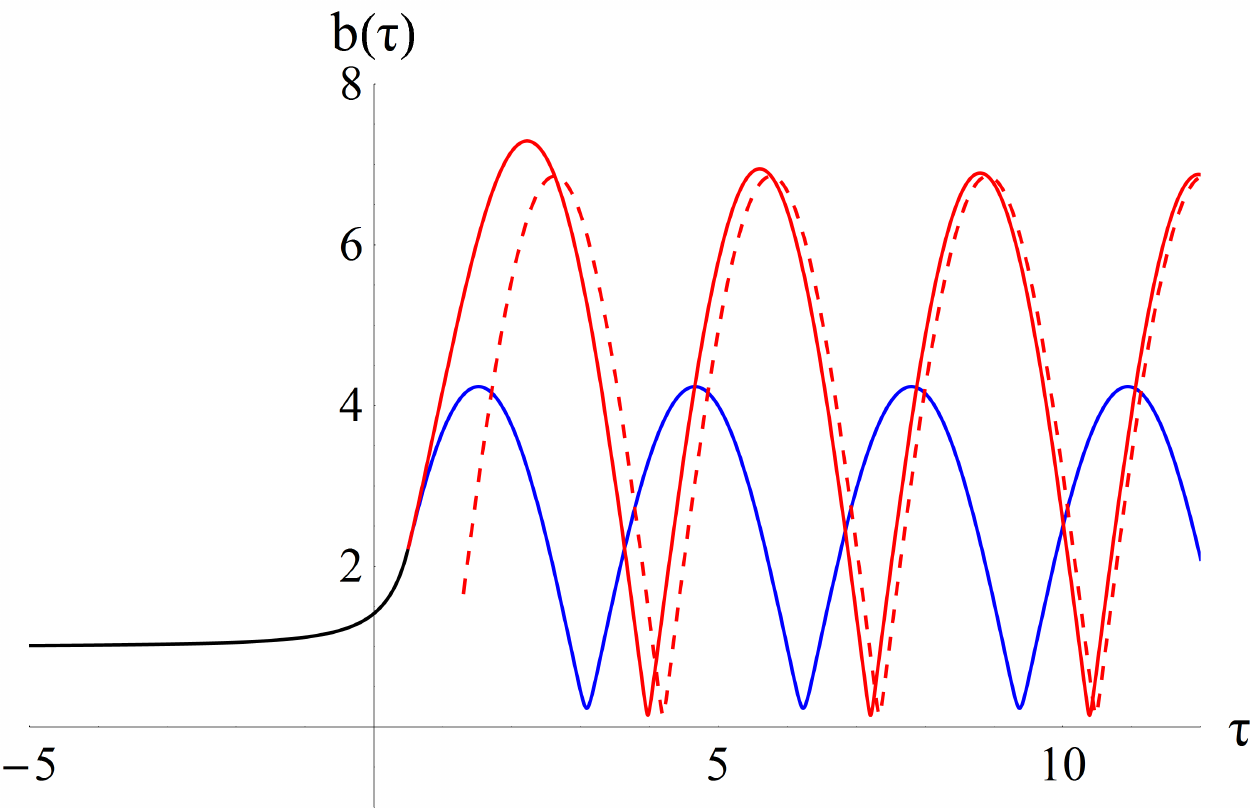} 
\includegraphics[width=0.45\textwidth]{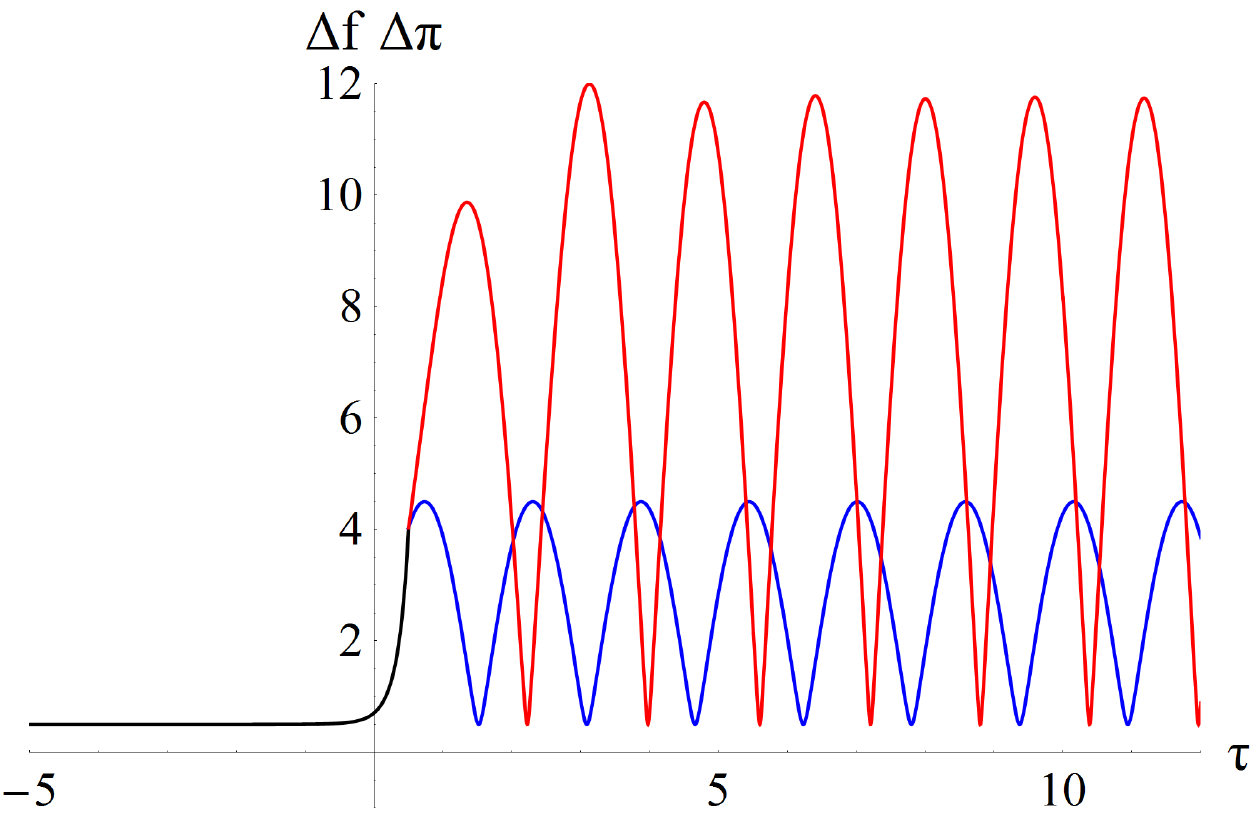} 
\caption{Left plot:
The form of the function $b(\tau)$ for $\omega_0=1$ and $H=2$, and $\tau_0=0.5$.
The black line corresponds to eq. (\ref{bdS2sf}), the blue line to eq. (\ref{bRD2f}),
the red solid line to eq. (\ref{bmd}) and the red dashed line to eq. (\ref{bMD2fas}).
Right plot: The product of uncertainties $\Delta f \Delta \pi$ during the evolution of
the wave function.
}
\label{fig3}
\end{figure}

In the left plot of figure \ref{fig3} we depict the evolution of the function $b(\tau)$ 
for the dS-RD scenario (black and blue lines) and the dS-MD scenario (black and red lines). 
We also depict the approximate expression (\ref{bMD2fas}) as a red dashed line. It 
captures well the exact form of $b(\tau)$ in the MD era for sufficiently large $\tau$.
The strong oscillatory form of $b(\tau)$ during RD and MD, with a frequency equal to $2\omega_0$,
is also apparent. 

In the right plot of figure \ref{fig3} we depict the product of uncertainties 
in $f$ and the conjugate momentum $\pi$, which, according to eqs. (\ref{varfvarp}), is given by
\begin{equation}
\Delta f \,\Delta \pi=\frac{1}{2}\sqrt{1+\frac{b^2(\tau)b'^2(\tau)}{\omega^2_0}}.
\label{uncertainty} \end{equation}
At early times this product is equal to 1/2, as the oscillator wave function is reduced to 
that of the ground state in a static background. A strong increase is observed during the
dS era, followed by strong oscillatory behaviour in both the RD and MD subsequent periods.
The minimum in each oscillation is 1/2. For $H \ll \omega_0$, we have $b(\tau)\simeq 1$ at
all times, so that $\Delta f \, \Delta \pi \simeq 1/2$. 
For $H \gg \omega_0$, the oscillations are pronounced and the 
maximum is of the order of a power of the ratio $H/\omega_0$. 
We have $\Delta f \, \Delta \pi \sim H/\omega_0$ at the end of inflation. During RD the
maximum of the oscillations is $\sim (H/\omega_0)^2$, while during MD it is $\sim (H/\omega_0)^3$.

\begin{figure}[t!]
\centering
\includegraphics[width=0.48\textwidth]{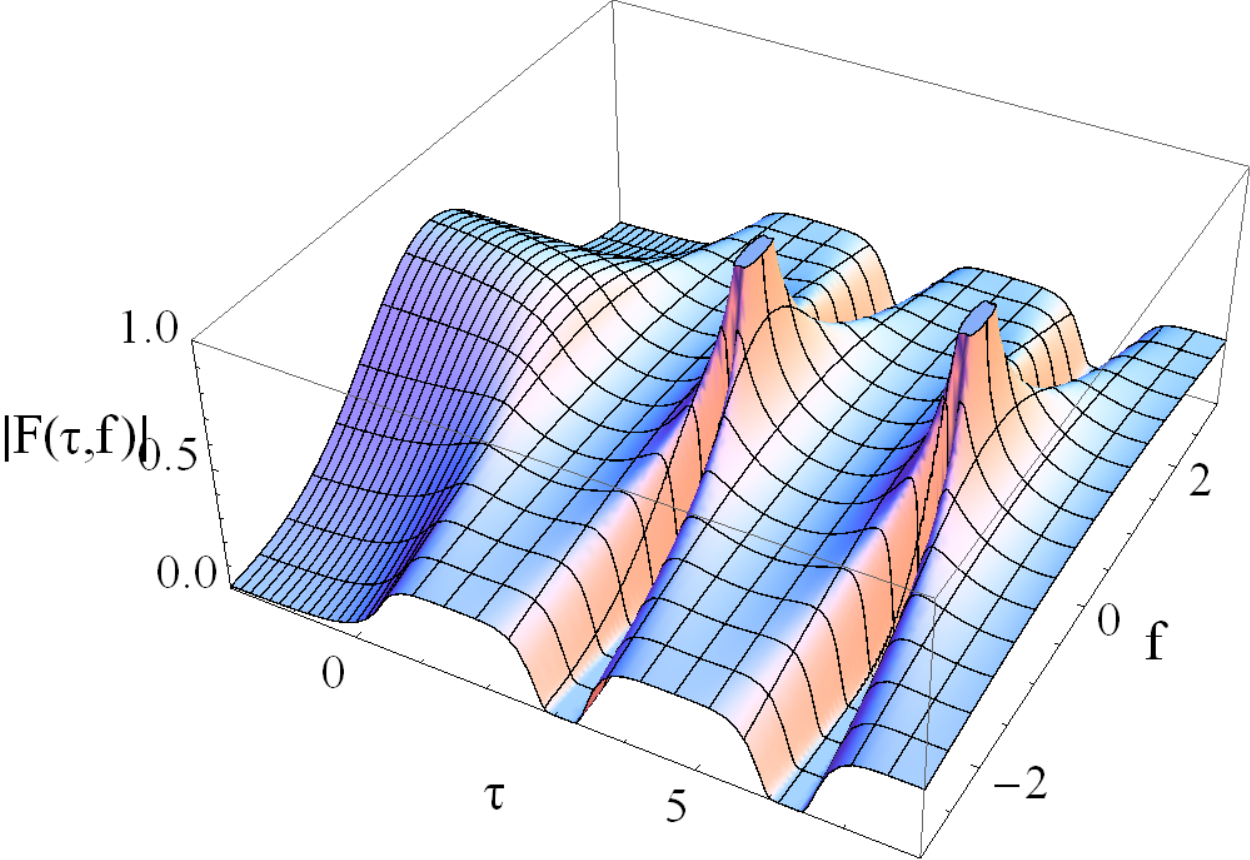} 
\includegraphics[width=0.48\textwidth]{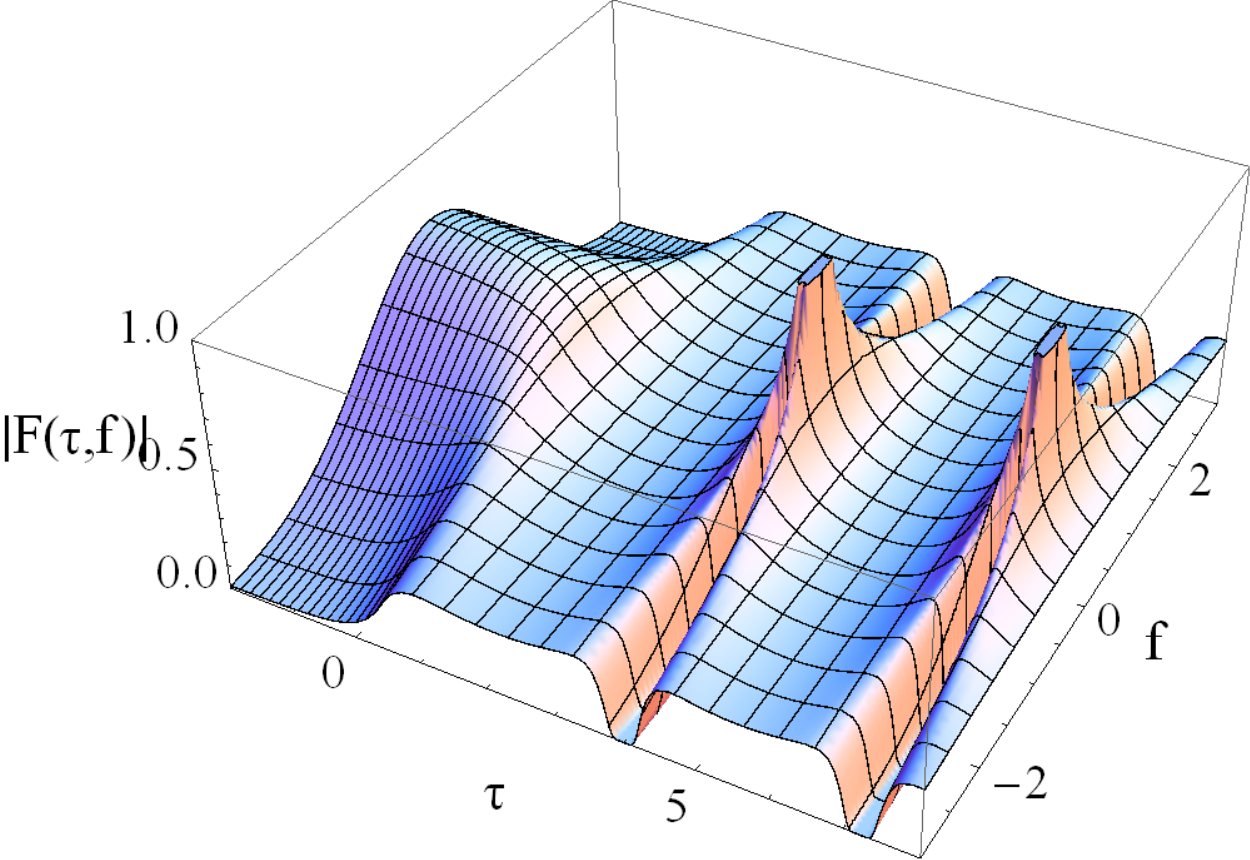} 
\caption{Left plot:
The amplitude of the `ground-state' wave function for the transition from
a dS to a RD background at $\tau=0.5$, for $\omega_0=1$, $H=2$. 
Right plot: Similarly to the left plot for the transition from a dS to a MD background.
}
\label{fig4}
\end{figure}

The form of the function $b(\tau)$ affects drastically the wave function of the quantum 
oscillator in the respective background. In figure \ref{fig4} we depict the amplitude of the 
wave function for a transition from a dS to a RD background (left plot) and from a dS to a 
MD background (right plot). The transition time has been chosen arbitrarily 
at $\tau_0=0.5$ and can be identified by the 
change of the grid density on the surface.
 The squeezing of the wave function originating in the dS regime
persists in the subsequent eras. Moreover, an oscillatory pattern appears with a period
set by $2\omega_0$, 
consistently with eqs. (\ref{bRD2f}) and (\ref{bMD2fas}).
Even though we do not depict them,
the real and imaginary parts of the wave function display strong oscillatory patterns along the 
$f$-direction during the time intervals that $b(\tau)$ varies strongly.
The general conclusion is that the non-trivial properties of the quantum state that 
develop during the dS era persist during the subsequent RD or MD era, while they are
supplemented by an oscillatory pattern in the amplitude with frequency twice the fundamental 
frequency of the oscillator.

\section{Entanglement entropy of two quantum oscillators}\label{EntEnt}

Significant technical difficulties in the calculation of the eigenvalues 
of the reduced density matrix
make the calculation of the entanglement entropy 
for the full field-theoretical problem a complicated process, which often lacks transparency. 
For this reason, 
we present here a complete analysis of the toy model of two coupled oscillators. 
Despite being a crude analogue of the full problem, we shall
see that this model displays surprising rich features. In particular, 
it demonstrates that IR 
contributions to the entropy can be significant, thus pointing towards effects  in the 
field-theoretical setup that may go beyond the UV divergent surface term $\sim R^2/\epsilon^2$.

We consider the entanglement between two coupled harmonic oscillators in an expanding background.
We assume that at very early times the background is inflationary and the coupled system is 
in the ground state of the standard harmonic oscillator. This is a reduction of the 
field-theoretical Bunch-Davies vacuum to the current simpler problem.
The system can be described through
the Hamiltonian
\begin{equation}  H=\frac{1}{2}\left[p_1^2+p_2^2+k_0(x_1^2+x_2^2)+k_1(x_1-x_2)^2-\lambda(\tau)
(x_1^2+x_2^2)\right].
\label{hamiltonian2osc}
\end{equation}
We have followed the notation of \cite{srednicki} and 
switched from the variable $f$ used in the previous sections, to the more standard variable $x$ for 
the harmonic oscillator.
The Hamiltonian can be rewritten as
\begin{equation}
    H=\frac{1}{2}\left[p_+^2+p_-^2+w_+^2(\tau)x_+^2+w_-^2(\tau)x_-^2\right],
\end{equation}
where $x_\pm=(x_1\pm x_2)/\sqrt{2}$ and
\begin{equation}
    w^2_\pm(\tau)=\omega_{0\pm}^2-\lambda(\tau),
\label{ompmt} \end{equation}
with $\omega_{0+}^2=k_0, \omega_{0-}^2=k_0+2k_1$.
In this way, the system can described in terms of two decoupled modes. 
For each one of them, the solution of the Schr\"odinger equation is given by 
eq. (\ref{solschrod}) in terms of the solution (\ref{quantharm}) of the standard
harmonic oscillator with constant frequency. For oscillators arising from 
a massive field in a dS background with a scale factor given by eq. (\ref{atau}), 
we have $\lambda(\tau)=2\kappa/\tau^2$ and $b(\tau)$ given by eq. (\ref{bsolfin}).
For a scale factor given by the 
expressions (\ref{scalefactors}) during the various eras,
and for oscillators corresponding to a massless field, 
we have $\lambda(\tau)=a''/a$. The function $b(\tau)$ is given by 
eq. (\ref{bdS2sf}) for a dS background, 
 by eq. (\ref{bRD2f}) during a subsequent RD period, 
 or by eq. (\ref{bmd}), which is well approximated at late times by eq. (\ref{bMD2fas}),
for a subsequent MD period.

The solution that reduces to the 
ground state $(n=0)$ of the standard harmonic oscillator at early times
is given by
\begin{equation}
    F_{\text{0}}(\tau,x)=\frac{1}{\sqrt{b(\tau)}}\left(\frac{\omega_0}{\pi}\right)^{1/4}
    \exp\left(-\frac{1}{2}\frac{\omega_0}{b^2(\tau)}x^2\right)\exp\left[\frac{i}{2}\left(\frac{b'(\tau)}{b(\tau)}x^2-\int d\tau \frac{\omega_0}{b^2(\tau)}\right)\right].
    \label{standardwavef}
\end{equation}
The phase containing the integral $\int d\tau/b^2(\tau)$ will be omitted in the following, 
as it does not affect any observables. 
On the contrary, this is not possible for the phase proportional to $b'/b$, which has measurable
effects.

The `ground state' of our toy model is the tensor product of the `ground states' of 
the two decoupled normal modes, given by 
\begin{equation}\psi_0(x_+,x_-)=\left(\frac{\Omega_+\Omega_-}{\pi^2}\right)^{1/4}\exp\left[-\frac{1}{2}\left(\Omega_+x_+^2+\Omega_-x_-^2\right)\right]\exp\left[\frac{i}{2}\left(G_+x_+^2+G_-x_-^2\right)\right],
\end{equation}
with
\begin{equation}
    \Omega_\pm(\tau)\equiv\frac{\omega_{0\pm}}{b^2(\tau;\omega_{0\pm})},\;\;\; G_{\pm}(\tau)\equiv\frac{b'(\tau;\omega_{0\pm})}{b(\tau;\omega_{0\pm})}.
    \label{OpmGpm}
\end{equation}
The wave function can be expressed in terms of the initial degrees of freedom $x_1$, $x_2$ as
\begin{multline}
\psi_0(x_1,x_2)=\left(\frac{\Omega_+\Omega_-}{\pi^2}\right)^{1/4}\exp\left[-\frac{1}{2}\left(\frac{\Omega_++\Omega_-}{2}(x_1^2+x_2^2)+(\Omega_+-\Omega_-)x_1x_2\right)\right]\\
    \times\exp\left[\frac{i}{2}\left(\frac{G_++G_-}{2}(x_1^2+x_2^2)+(G_+-G_-)x_1x_2\right)\right].
\end{multline}

We are now able to trace out one of the two degrees of freedom, in order to calculate the entanglement entropy. The reduced density matrix is given by
\be
\rho(x_2,x_2')=\int_{-\infty}^{+\infty}dx_1\psi_0(x_1,x_2)\psi_0^*(x_1,x_2').
\label{desint} \ee
By completing the square for the Gaussian integral and carrying out the integration
we find 
\be
\rho(x_2,x_2')=\pi^{-1/2}\left(\gamma-\beta\right)^{1/2}\exp\left(-\frac{\gamma}{2}(x_2^2+x_2'^2)+\beta x_2x_2'\right)
\exp\left( i\frac{\delta}{2}(x_2^2-x_2'^2)\right),
\label{densres} \ee
where 
\begin{eqnarray}
    \gamma&=&\frac{\Omega_++\Omega_-}{2}-\frac{(\Omega_+-\Omega_-)^2-(G_+-G_-)^2}{4(\Omega_++\Omega_-)},
\label{gammar} \\
    \beta&=&\frac{(\Omega_+-\Omega_-)^2+(G_+-G_-)^2}{4(\Omega_++\Omega_-)},
\label{betar} \\
 \delta&=&\frac{G_++G_-}{2}-\frac{(\Omega_+-\Omega_-)(G_+-G_-)}{2(\Omega_++\Omega_-)}.
\label{deltar}
\end{eqnarray}
 
The calculation of the entanglement entropy requires the determination of the 
eigenfunctions $f_n(x_2)$ and eigenvalues $p_n$ of the reduced density matrix, 
which satisfy
\be
 \int_{-\infty}^{+\infty}dx_2'\rho(x_2,x_2')f_n(x_2')=p_n f_n(x_2).
 \label{eigenfv} \ee
The eigenfunctions are simple generalizations of those of the original problem for a static
background with $\delta=0$ \cite{srednicki}. They are 
\begin{equation}
    f_n(x)=H_n(\sqrt{\alpha}x)\exp\left(-\frac{\alpha}{2} x^2\right)\exp\left( i \frac{\delta}{2} x^2\right),
\label{eigennn} \end{equation}
where 
\be
\alpha=\sqrt{\gamma^2-\beta^2}
\label{asq} \ee
and $H_n$ is the Hermite polynomial of order $n$. A simple way to check this result is
to express the Hermite polynomial 
in terms of the generating function
\be
g(t,x)=\exp \left(2a x t-c^2 t^2 \right)
\label{hermgen} \ee
by means of the relation
\be
c^n H_n\left( \frac{a}{c} x \right)=\left. \frac{\partial^n g(t,x)}{\partial t^n} \right|_{t=0}.
\label{hermitee} \ee
Replacing $H_n(\sqrt{\alpha}x)$ in eq. (\ref{eigennn}) through the above expression with $a=\sqrt{\alpha}$
and $c=1$, and substituting into eq. (\ref{eigenfv}), gives
\begin{multline}
 \int_{-\infty}^{+\infty} dx_2'\rho(x_2,x_2')f_n(x_2')\\
 =
\sqrt{\frac{2(\gamma-\beta)}{\gamma+\alpha}}
\left. \frac{\partial^n }{\partial t^n} 
\exp \left(\frac{2\sqrt{\alpha}\beta}{\alpha+\gamma}t x_2-\frac{\gamma-\alpha}{\gamma+\alpha}t^2
+\frac{\beta^2-\gamma^2-\alpha \gamma}{2(\alpha+\gamma)}x_2^2 \right)
\exp\left( i\frac{\delta}{2}x_2^2\right)
  \right|_{t=0}.
\label{solgener}
\end{multline}
By imposing the relation (\ref{asq}) and making use of eqs. (\ref{hermgen}), (\ref{hermitee}) with the values
$c=\sqrt{(\gamma-\alpha)/(\gamma+\alpha)}$ and $a=\sqrt{\alpha}\beta/(\alpha+\gamma)=\sqrt{\alpha}c$, 
it can be
easily seen that the above expression becomes proportional to the eigenfunction (\ref{eigennn}).
The multiplicative factor fixes 
the eigenvalues $p_n$, which are given by the same expression as in \cite{srednicki}:
\begin{equation}
    p_n=\sqrt{\frac{2(\gamma-\beta)}{\gamma+\alpha}}\left(\frac{\beta}{\gamma+\alpha}\right)^n=(1-\xi)\xi^n,
    \label{pn}
\end{equation}
where
\begin{equation}
    \xi=\frac{\beta}{\gamma+\alpha}.
    \label{xiba}
\end{equation}
Notice that they satisfy
\begin{equation}
    \sum_{n=0}^{\infty}p_n=(1-\xi)\sum_{n=0}^{\infty}\xi^n= 1.
    \label{pnsum}
\end{equation}
Finally, the entanglement entropy can be calculated as
\begin{equation}
    S=-\sum_{n=0}^{\infty}(1-\xi)\xi^n\ln\left[(1-\xi)\xi^n\right]=
    -\ln\left(1-\xi\right)-\frac{\xi}{1-\xi}\ln\xi.
\label{ent2osc}
\end{equation}

\subsection{De Sitter background}

\begin{figure}[t!]
\centering
\includegraphics[width=0.48\textwidth]{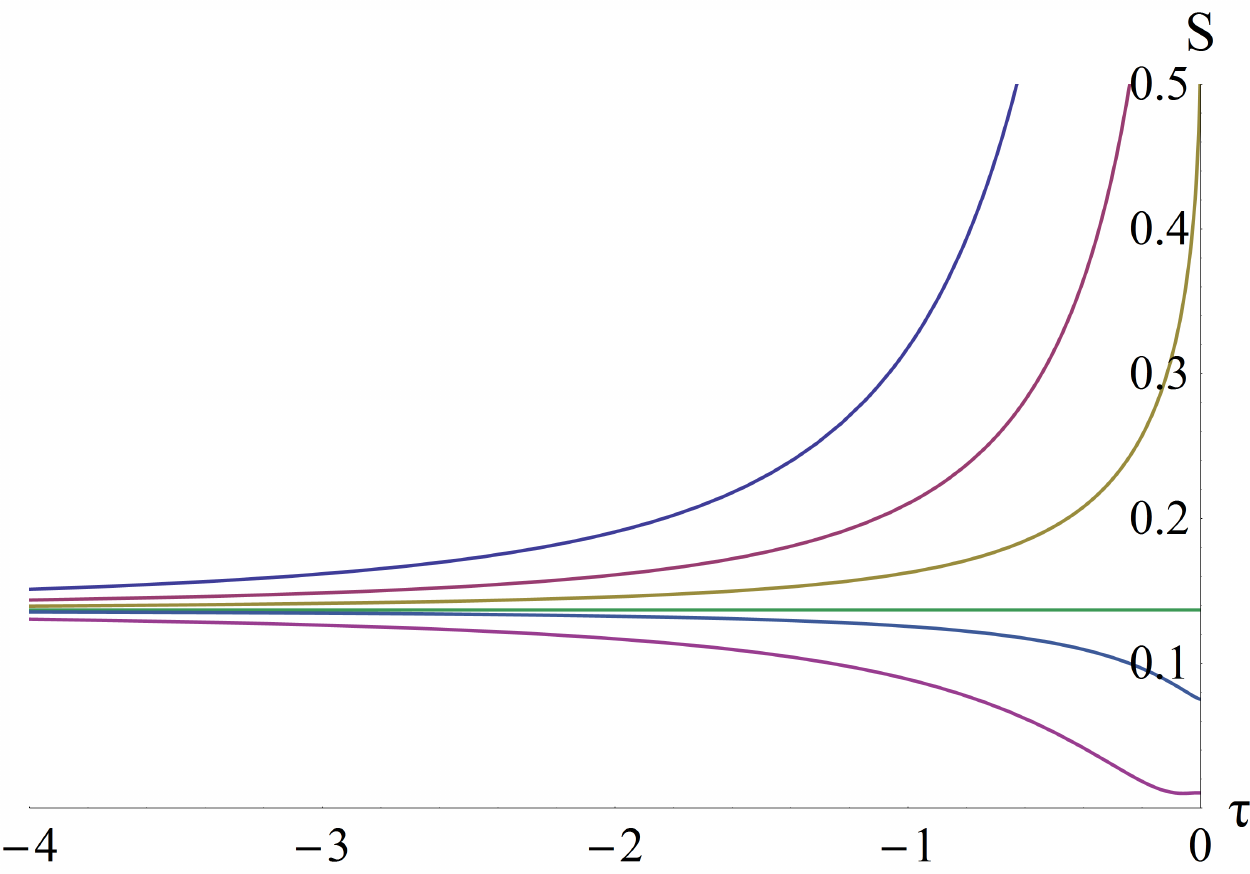} 
\includegraphics[width=0.48\textwidth]{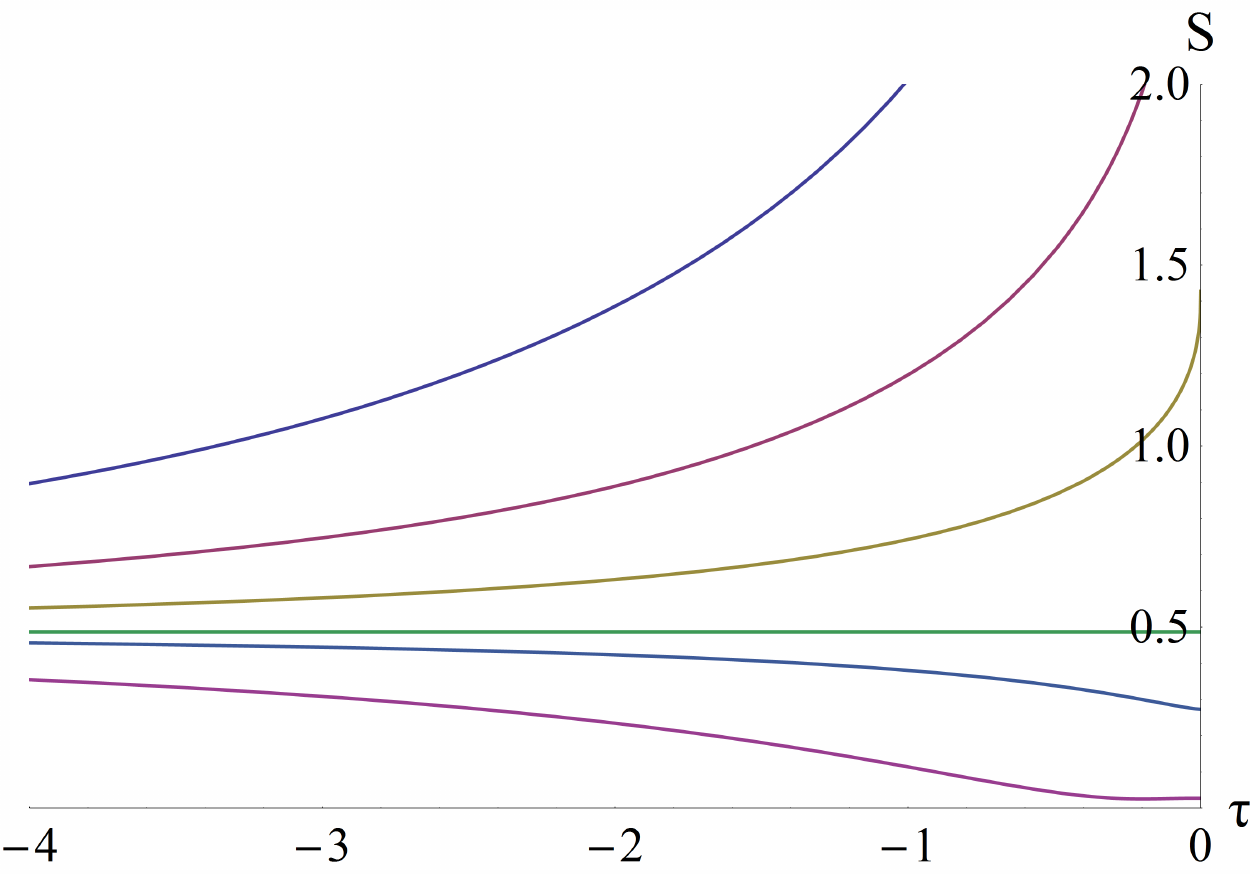} 
\caption{
Left plot: The entanglement entropy in a dS background as a function of conformal time $\tau$ 
for $\omega_+=1$, $\omega_-=2$ and $\kappa=$1, 0.5, 0.2, 0, $-0.1$, $-0.5$ (from top to bottom). 
Right plot: Similarly to the left plot for $\omega_+=1$, $\omega_-=0.2$.
}
\label{fig7}
\end{figure}

The entanglement entropy (\ref{ent2osc})
in a dS background is depicted in figure \ref{fig7} for various values of the
parameter $\kappa$, which can be linked to the 
mass of a free field in this background through eq. (\ref{kappa}). In this context,
the two oscillators would correspond to values of the quantum field at two points is space,
which would then be coupled through the spatial derivative term of the action (\ref{action1}). 
Clearly, the toy model we are considering is a very coarse reduction of the original problem of
interest, which concerns the entanglement entropy of a quantum field \cite{srednicki}. 
However, it reveals some fundamental features of the expected form of the entanglement
entropy, such as its dependence on the conformal time. 
The background in figure \ref{fig7} corresponds to exact dS space, with the conformal 
time $\tau$ varying between $-\infty$ and 0, and the scale factor varying between 0 and $\infty$.
(In the parametrization of the first of eqs. (\ref{scalefactors}), one must set $\tau_0=-1/H$ in this case.)

In the left plot of figure \ref{fig7},
the fundamental frequencies of the two decoupled oscillators have been taken 
as $\omega_+=1$ and $\omega_-=2$
in arbitrary units, which would correspond to $k_0=1$ and $k_2=1.5$ for the parameters of the
Hamiltonian (\ref{hamiltonian2osc}). 
Notice that for an exact dS background, parametrized through the metric (\ref{dsmetric}) 
in terms of conformal time, the evolution of the wave function of each normal mode depends 
only on the dimensionless combination $\omega\tau$ and the dimensionless parameter $\kappa$.
An absolute scale can be set by selecting the time $\tau$ 
at which the scale factor $a(\tau)=-1/(H\tau)$ becomes equal to 1.
In the right plot of figure \ref{fig7}, the frequency of the second decoupled oscillator has been
taken equal to $\omega_-=0.2$, which would correspond to $k_0=1$ and $k_2=-0.48$.
Notice the difference in the scale of the vertical axis in the two plots. A smaller value of
$\omega_-$ implies that the wave function of the corresponding oscillator gets squeezed at an earlier time through the analogue of the process of `horizon exit'. As a result, 
entanglement becomes stronger earlier. 

The top line in each of the plots of
figure \ref{fig7} corresponds to the entanglement entropy for $\kappa=1$, the value
that, according to eq. (\ref{kappa}), would arise for a massless field. A strong increase of the entanglement
entropy occurs as $\tau\to 0^-$.
The straight horizontal line corresponds to a conformally coupled field with $m^2=2H^2$ and 
$\kappa=0$. In this case the entropy remains constant and equal to its value in 
a static background. Intermediate values of $\kappa$ generate the curves between these two
cases. Negative values of $\kappa$, which would arise for massive fields with $m^2>2H^2$, 
result in the reduction of the entropy at late times. Clearly, such massive excitations 
become suppressed in the dS background and this is reflected in the entropy.

For the massless case with $\kappa=1$, in the limit $\tau\to -\infty$ we have $b(\tau)\to 1$ and
\be
\xi\simeq \left(\frac{\sqrt{\omega_-}-\sqrt{\omega_+}}{\sqrt{\omega_-}+\sqrt{\omega_+}} \right)^2.
\label{xiinf} \ee
The entanglement entropy, given by eq. (\ref{ent2osc}), reproduces the result for 
a static background \cite{srednicki}.

The entropy vanishes at all times if $\omega_+=\omega_-$ i.e. if the two
oscillators are decoupled. 
However, for $\omega_+\not= \omega_-$, in the limit $\tau\to 0^-$ we have 
\be
b(\tau)\simeq -\frac{1}{\omega_0\tau}-\frac{\omega_0 \tau}{2},
 \quad \quad
\frac{b'(\tau)}{b(\tau)}\simeq -\frac{1}{\tau}+\omega_0^2\tau
\label{bbpappr} \ee
and 
\be
\xi \simeq 1- \left|\frac{4\, (\omega_+\omega_-)^{3/2}}{\omega^2_- - \omega_+^2} \tau \right|.
\label{xi0} \ee
The entanglement entropy can be approximated as
\be
S_{\rm dS} \simeq 1- \ln \left|\frac{4\, (\omega_+\omega_-)^{3/2}}{\omega^2_- - \omega_+^2} \tau \right|. 
\label{entapprdS} \ee
If inflation lasts for a long time, the entanglement entropy increases linearly with the number 
of efoldings ${\cal N}=\ln a(\tau)= -\ln|H\tau|$. 
We can express this result in terms of the physical (redshifted) 
frequencies $\omegat_\pm=\omega_\pm/a(\tau)$ as 
\be
S_{\rm dS} \simeq 1- \ln \left|\frac{4\, (\omegat_+\omegat_-)^{3/2}}{\omegat^2_- - \omegat_+^2} \frac{1}{H} \right|.
\label{entapprdSr} \ee
From this point of view, the entanglement becomes significant when the physical frequencies drop much
below the Hubble scale. 
For oscillators associated with the modes of a massless field with Hamiltonian
given by eq. (\ref{eq:Hamiltonian_discretized}), this is equivalent to the respective wavelength 
growing beyond the Hubble radius (`horizon crossing') and the modes freezing. 
Our analysis reveals the source of this
behaviour: the continuous squeezing of the mode wave function.

The direct comparison of eqs. (\ref{xiinf}), (\ref{xi0}) is also illuminating. 
If one uses the physical frequencies $\omegat_\pm$, which become very high for $\tau\to-\infty$, 
eq. (\ref{xiinf}) retains
its form, without the appearance of the Hubble scale $H$. 
This implies that the contribution to the entanglement entropy 
from high-frequency modes of a complex system, such as a quantum field, 
is expected to be similar to a static background 
for such modes, even if they are redshifted by
the expansion. On the other hand, modes of the system that are low-frequency relative
to the Hubble scale, give large contributions to the entanglement entropy.

\subsection{Radiation and matter domination}

\begin{figure}[t!]
\centering
\includegraphics[width=0.45\textwidth]{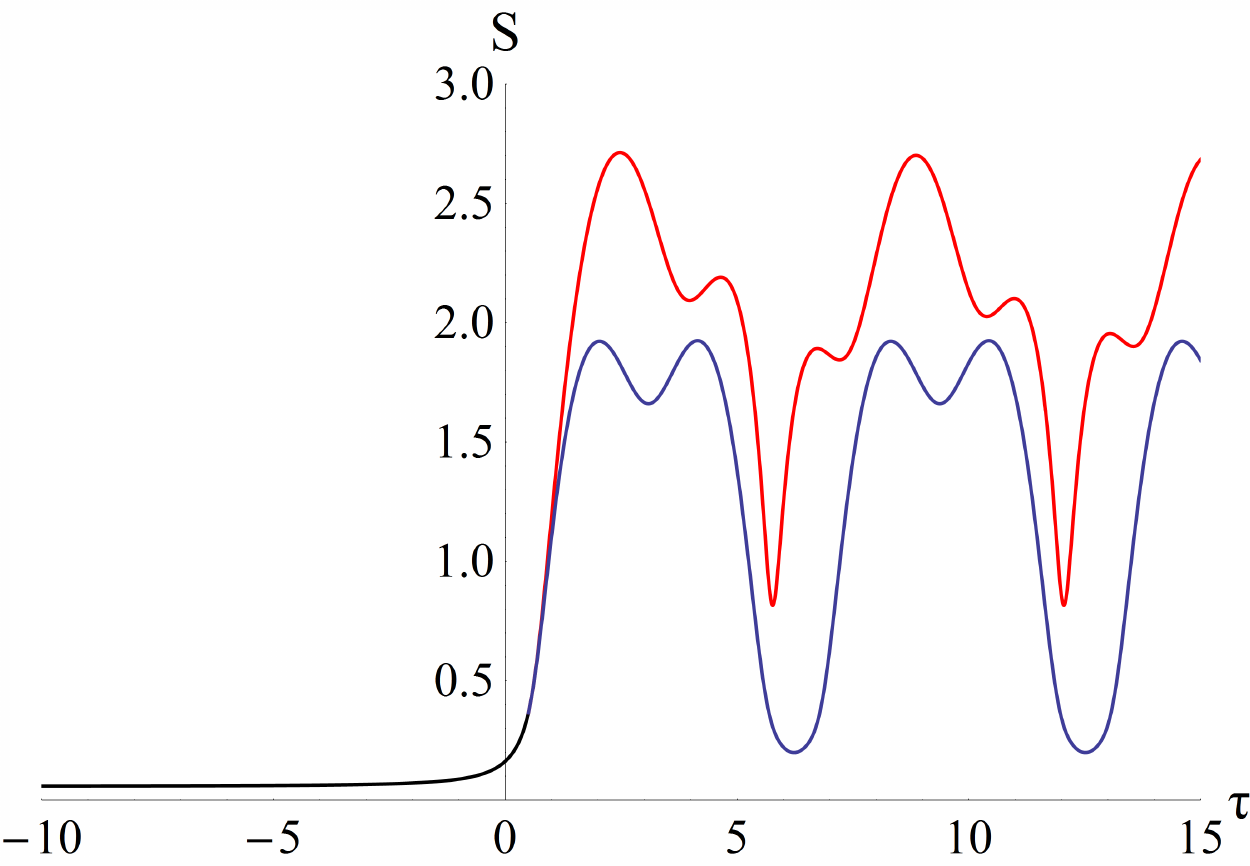} 
\includegraphics[width=0.45\textwidth]{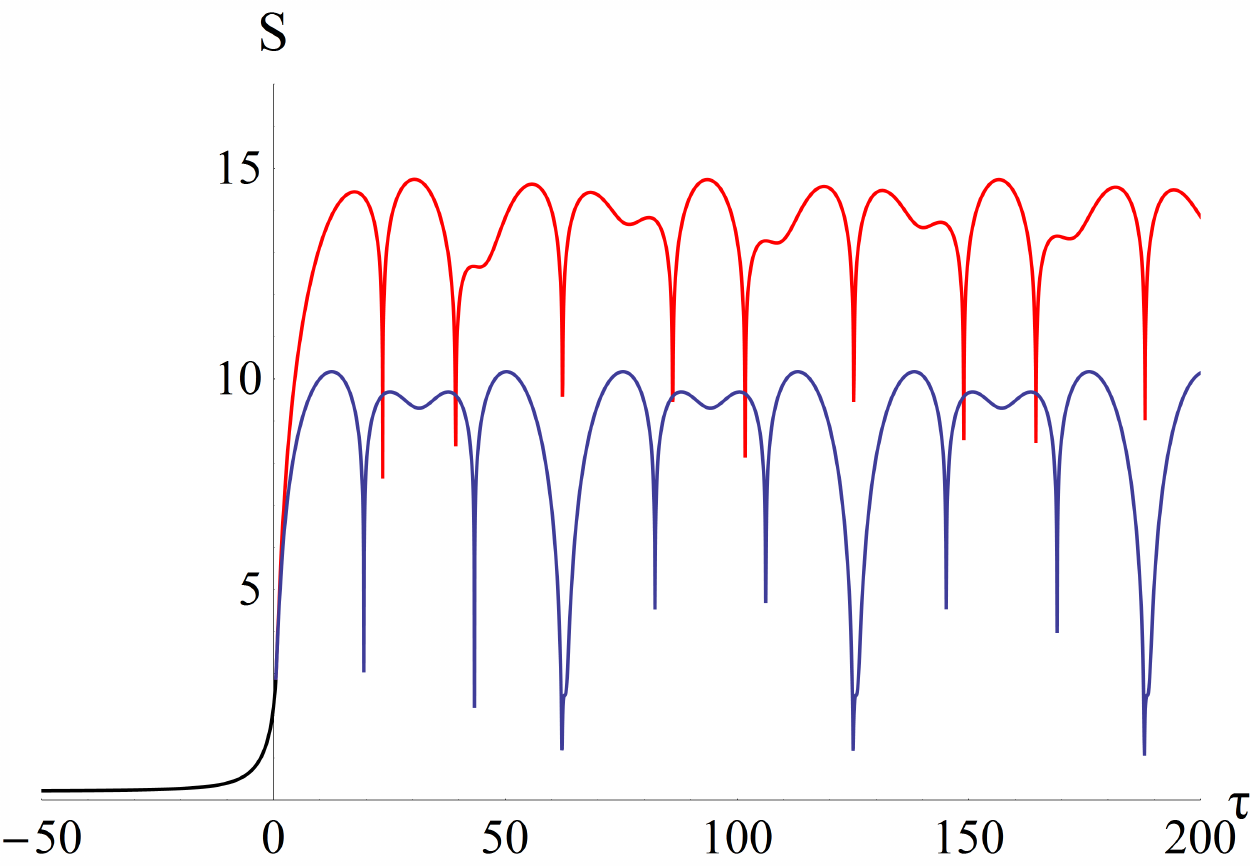}
\caption{
Left plot: The entanglement entropy as a function of conformal time $\tau$
for $\omega_+=1$, $\omega_-=1.5$, $H=2$ and $\tau_0=0.5$.
The black line corresponds to a dS background, with a transition at $\tau_0$ either to 
a RD era (blue line) or to a MD era (red line). Right plot: Similarly to the left plot 
for $\omega_+=0.1$, $\omega_-=0.25$, $H=2$ and $\tau_0=0.5$.
}
\label{fig8}
\end{figure}

A crucial question is whether the entanglement entropy developed during the 
dS era survives during a subsequent RD or MD era. At very early times we have
$\omega_\pm |\tau| =(\omega_\pm/a(\tau))/H \gg 1$ for both oscillators. This can be interpreted as
the corresponding modes being `subhorizon'. During the dS evolution a `horizon crossing' occurs at 
a time $\tau_c$ given by eq. (\ref{exit}) for each oscillator. The strong entanglement effects 
appear after such a time. During the subsequent RD or MD era, the reduction of the product $a(\tau)H(\tau)$ 
implies that at some sufficiently late time the condition $\omega_\pm |\tau| =(\omega_\pm/a(\tau))/H \gg 1$
will be satisfied again for both oscillators, which will have `crossed into the horizon' again. 
In the standard approach to cosmology, modes that have re-entered the horizon are treated as
classical stochastic variables, according to the discussion in section \ref{QMS}. 
It is very interesting to understand how a purely quantum property of the system, such as the
entanglement entropy, would behave in this regime. 

In figure \ref{fig8} we depict the evolution of the entanglement entropy for a transition 
from a dS era (black line) to a RD era (blue line) or a MD era (red line). The left plot
was generated for frequencies  $\omega_+=1$, $\omega_-=1.5$, while the right one for 
lower frequencies $\omega_+=0.1$, $\omega_-=0.25$, in order to observe the difference in
time scales and oscillatory patterns. It is apparent that the entanglement is not 
destroyed by the transition, but retains asymptotically a constant average value. 
This is not unexpected, as the wave functions preserve their
squeezed nature, which is modulated by oscillations resulting from a combination of the
fundamental frequencies $\omega_+$, $\omega_-$. 
The entropy drops within short intervals during the oscillations, but its
average value remains constant. It is noticeable that this value is larger than the
one at the end of inflation. The reason can be found by comparing the expressions (\ref{bdS2sf}),
(\ref{bRD2f}), (\ref{bMD2fas}) for the function $b(\tau)$ that determines the form of the
wave function during each era. For $H\gg \omega_0$, the functions 
$b_{\rm RD}(\tau)$ and $b_{\rm MD}(\tau)$ are 
enhanced by additional powers of $H/\omega_0$ relative to $b_{\rm dS}(\tau)$.

\section{The implications for the quantum field} \label{QF}

\subsection{The reduced density matrix}

The generalization from the case of two oscillators to a system of $N$ coupled oscillators, with a
Hamiltonian such as (\ref{eq:Hamiltonian_discretized}) arising
from the discretization of the field-theoretical problem, proceeds along the lines of \cite{srednicki}.
If the normal modes of the system lie in a squeezed state with a wave function given by 
eq. (\ref{standardwavef}), the reduced density matrix is given by a generalization of 
eq. (\ref{densres}). When $n$ oscillators are traced out, it has the form 
\begin{multline}
\rho(\bx_2,\bx'_2)=\left( \frac{ \det {\rm Re} (\gamma-\beta)}{\pi^{N-n}} \right)^{1/2} \\
\times \exp \left( -\frac{1}{2}\bx_2^T\gamma\, \bx_2  -\frac{1}{2}\bx^{\prime T}_2\gamma\, \bx'_2 +\bx^{\prime T}_2 \beta\, \bx_2
+\frac{i}{2}\bx_2^T\delta\, \bx_2  -\frac{i}{2}\bx^{\prime T}_2\delta \, \bx'_2
\right).
\label{densresN} \end{multline}
Here $\gamma$ and $\delta$ are $(N-n)\times(N-n)$ real symmetric matrices, while $\beta$ is a 
$(N-n)\times(N-n)$ Hermitian matrix. Similarly to the case of two oscillators, it can be shown that
the eigenvalues of the density matrix do not depend on $\delta$. A major technical difficulty arises
because the matrices
$\gamma$ and $\beta$ cannot be diagonalized through real orthogonal transformations in order to 
identify the eigenvalues of the reduced density matrix. 
These are guaranteed to be real by the nature of the density matrix, but
the determination of their exact values requires a separate extensive analysis, which is 
presented in \cite{preparation}. Notice that, in the case of 
two oscillators, $\beta$ is reduced to a single real parameter, a simplification that made the calculation of the 
spectrum of the density matrix in the previous section feasible. 

Even though the complete analysis of the 
field-theoretical case in $3+1$ dimensions
will be deferred to future work, it is still possible to draw a 
qualitative conclusion about the form of the entanglement entropy during the dS era.
For this, we need to consider how eq. (\ref{densresN}) arises.
For a system of $N$ coupled quantum harmonic oscillators, such as those described by the 
Hamiltonian (\ref{eq:Hamiltonian_discretized}), 
there is an orthogonal transformation $O$ relating the coordinates $x_k$ (we use $x_k$ instead
of the notation $f_{lm,k}$ that we use in section \ref{subsec:exp_coordinate}) to the normal coordinates $\tilde{x}_k$, which reduces the system to a set of 
decoupled harmonic oscillators, one for each normal mode. 
We use the compact notation
\begin{equation}
\mathbf{x} = \left( \begin{array}{c} x_1 \\ x_2 \\ \vdots \\ x_N \end{array} \right) , 
\quad \quad
\mathbf{\xt} = \left( \begin{array}{c} \xt_1 \\ \xt_2 \\ \vdots \\ \xt_N \end{array} \right) , 
\quad \quad
\mathbf{\tilde{x}} = O \mathbf{x} .
\end{equation}
We consider states of the system where all normal modes lie in a squeezed state
of the form (\ref{standardwavef}). At any given time, the total wave function is
\begin{equation}
\Psi \left( \mathbf{\tilde{x}} \right)  \sim 
 \exp \left( - \frac{1}{2}  \mathbf{\tilde{x}}^T \tilde{\Omega}\, 
 \mathbf{\tilde{x}}  \right) ,
 \label{Omegatilde}
\end{equation}
where $\tilde{\Omega}$ is the diagonal matrix whose diagonal elements are the complex quantities 
\begin{equation}
\tilde{\Omega}_{k} = 
\frac{\omega_k}{b^2(\tau,\omega_k)}-i\frac{b'(\tau,\omega_k)}{b(\tau,\omega_k)} ,
\label{Omegai}
\end{equation}
with the prime denoting a derivative with respect to $\tau$.
In the last equation
we indicated explicitly the dependence of the quantity $b$ on the eigenfrequency 
$\omega_k$ of each normal mode.

We define as subsystem $1$ the set of $n$ oscillators described by the coordinates $x_j$, where $j \leq n$. The $N - n$ oscillators described by coordinates $x_j$ 
with $j > n$ comprise the complementary subsystem, which we call subsystem 2. 
We would like to trace out one subsystem in order to find the reduced density matrix 
for the other. Because of the fundamental properties of entanglement,  
tracing out sybsystem 1 leads to the same value for the entropy as tracing out subsystem 2, 
as long as the total system lies in a pure state. We follow the convention of \cite{srednicki} and trace out subsystem 1.
For this purpose, we need to express the state in terms of the original coordinates $\mathbf{x}$,
\begin{equation}
\Psi \left( \mathbf{{x}} \right)  \sim 
 \exp \left( - \frac{1}{2}  \mathbf{{x}}^T {\Omega} \,
 \mathbf{{x}}  \right)  ,
 \label{Omega}
\end{equation}
where $\Omega = O^T \tilde{\Omega}\, O$. 
The matrix $\Omega$ is a complex symmetric matrix. 
The density matrix describing the overall system assumes the form
\begin{equation}
\rho \left( \mathbf{x} ; \mathbf{x}^\prime \right) \sim
 \exp \left[ - \frac{1}{2}  \left( 
 \mathbf{{x}}^T {\Omega}\, \mathbf{{x}}  + \mathbf{{x}}^{\prime T} {\Omega}^*\, \mathbf{{x}}'  
 \right)  \right].
\end{equation}
We use the block-form notation \cite{srednicki}
\begin{equation}
\Omega = \left( \begin{array}{cc} A & B \\ B^T & C \end{array} \right) , \quad\quad \mathbf{x} = \left( \begin{array}{c} \mathbf{x}_1 \\ \mathbf{x}_2 \end{array} \right)  ,
\label{block-form}
\end{equation}
where the matrix $A$ is an $n \times n$ matrix, the matrix $C$ is an $\left( N - n \right) \times \left( N - n \right)$ matrix and so on. Notice that the matrices $A$ and $C$ are complex symmetric matrices, whereas the matrix $B$ is not a square matrix.

The reduced density matrix $\rho_2 \left( \mathbf{x}_2 ; \mathbf{x}_2^\prime \right) = \int d^n \mathbf{x}_1 \rho \left( \mathbf{x}_1 , \mathbf{x}_2 ; \mathbf{x}_1 , \mathbf{x}_2^\prime \right)$, which describes subsystem 2,
can be found by evaluating the multidimensional Gaussian integrals. 
The result, with the appropriate normalization constant, 
is given by eq. (\ref{densresN}) with
\begin{eqnarray}
\gamma -i \delta&=& C - \frac{1}{2} B^T {\rm Re} \left( A \right)^{-1} B , 
\label{gammam} \\
\beta &=& \frac{1}{2} B^\dagger {\rm Re} \left( A \right)^{-1} B .
\label{betam}\end{eqnarray}
It follows that $\gamma$ and $\delta$ are 
real symmetric matrices, whereas $\beta$ is a Hermitian matrix. 

As in the case of two oscillators that we analysed explicitly, the eigenvalues
do not depend on the matrix $\delta$.
An eigenfunction
 ${f} \left( \mathbf{x}_2 \right)$  of the density matrix 
 with eigenvalue $\lambda$ satisfies 
\begin{equation}
\int d^{N - n}\, \mathbf{x}_2^\prime\, {\rho}_2 \left( \mathbf{x}_2 ; \mathbf{x}_2^\prime
 \right) {f} \left( \mathbf{x}_2^\prime \right) = \lambda {f} \left( \mathbf{x}_2 \right) .
\end{equation} 
It can be seen easily that this eigenvalue is the same as that of the reduced density 
matrix $\tilde{\rho}_2 \left( \mathbf{x}_2 ; \mathbf{x}_2^\prime \right)$
resulting through setting $\delta=0$. The corresponding eigenfunctions are 
related through  
 $f \left( \mathbf{x}_2 \right) = \exp \left( - \frac{i}{2} \mathbf{x}_2^T 
  \delta\, \mathbf{x}_2 \right) \tilde{f} \left( \mathbf{x}_2 \right)$.
 Therefore, as long as we are interested in calculating the entanglement entropy, it suffices to find the spectrum of the matrix $\tilde{\rho}_2$. 

In the case the matrix $\beta$ is real, we know that the spectrum of the reduced density matrix is of the form \cite{srednicki}
\begin{equation}
p_{\left\{n_i\right\}} = \prod_{i=1}^M \left( 1 - \xi_i \right) \xi_i^{n_i} ,
\label{eq:reduced_spectrum}
\end{equation}
where $M=N-n$ is the number of the degrees of freedom that have not been traced out and
$n_i$ take values in the natural numbers.
The spectrum of the reduced density matrix is 
of the form \eqref{eq:reduced_spectrum} even when the matrix $\beta$ is not real. The 
proof is quite complicated, and the reader is referred to \cite{preparation} 
for all the details. Here we include a short summary of the basic points.

\subsection{The eigenvalues of the reduced density matrix}

In the case that the matrix $\beta$ is real, one can show that there is a series of three coordinate transformations that allow the factorization of the reduced density matrix \eqref{densresN} in terms of some kind of `canonical' coordinates. These are an orthogonal transformation that diagonalizes the matrix $\gamma$, a coordinate rescaling that 
sets the matrix $\gamma$ equal to the identity matrix, and finally an orthogonal transformation that diagonalizes $\beta$. 
This series of operations allows the writing of the reduced density matrix \eqref{densresN} as the tensor product of matrices of the form
\be
\rho(y_2,y_2')=\pi^{-1/2}\left(1-\beta\right)^{1/2}\exp\left(-\frac{1}{2}(y_2^2+y_2'^2)+\beta y_2\,y_2'\right),
\ee
describing \emph{one} `canonical' coordinate each. We have already seen in section \ref{EntEnt} that the eigenstates of this density matrix are the eigenstates of an effective simple harmonic oscillator with eigenfrequency equal to $\sqrt{1-\beta^2}$. Its spectrum is of the form $p_n = \left( 1 - \xi \right) \xi^n$, where $\xi = \beta/(1+\sqrt{1-\beta^2})$. 
The factorization of the reduced density matrix into a product of matrices of the above form directly implies that the eigenstates of the reduced density matrix are identical to the eigenstates of a system of $N - n$ \emph{effective} uncoupled simple harmonic oscillators, 
corresponding to the effective `canonical' normal modes, 
 and that its spectrum is of the form \eqref{eq:reduced_spectrum}.

When the matrix $\beta$ is not real, it is not possible to diagonalize it through 
the third transformation. 
The reduced density matrix cannot be factored to the tensor product of 
matrices describing a single degree of freedom each.
We can still perform the first two coordinate transformations and express the reduced density matrix in the form
\begin{equation}
\tilde{\rho}_2 \left( \tilde{\mathbf{x}}_2 ; \tilde{\mathbf{x}}_2^\prime \right) = \sqrt{\frac{\det \left( I - {\rm Re} \left(\tilde{\beta}\right) \right)}{\pi^{N - n}}} \exp \left[ - \frac{1}{2} \left( \tilde{\mathbf{y}}_2^T \tilde{\mathbf{y}}_2 + \tilde{\mathbf{y}}_2^{\prime T} \tilde{\mathbf{y}}_2^\prime \right) + \tilde{\mathbf{y}}_2^{\prime T} \tilde{\beta} \tilde{\mathbf{y}}_2 \right] ,
\end{equation}
where
\begin{equation}
\tilde{\beta} = \gamma^{-\frac{1}{2}} \beta \gamma^{-\frac{1}{2}} 
\label{betatildem}
\end{equation}
is now Hermitian.
It is remarkable that
the general structure of the eigenstates and eigenvalues 
of the reduced density matrix remains the same as when $\tilde{\beta}$ is real,
even though they are severely deformed (especially the eigenstates).

As in the case of a real $\tilde{\beta}$, there is a `ground' eigenstate of the form
\begin{equation}
\Psi_0 \left( \mathbf{x} \right) =c_0 \exp \left( - \frac{1}{2} \mathbf{x}^T {\cal W} \mathbf{x} \right) ,
\label{eq:eigenstate_ground}
\end{equation}
where the matrix ${\cal W}$ satisfies the quadratic equation
\begin{equation}
{\cal W} = I - \tilde{\beta}^T \left( I + {\cal W} \right)^{- 1} \tilde{\beta} .
\label{eq:omega}
\end{equation}
This equation has many solutions, but 
only one corresponds to a normalizable state, such as \eqref{eq:eigenstate_ground}.

In the case of a real matrix $\tilde{\beta}$, there would exist $N-n$ `first excited' eigenstates, one for each `canonical' mode of the reduced system. 
These states would be simply the product of the corresponding
`canonical' coordinate with the Gaussian `ground' eigenstate. In the case of complex $\tilde{\beta}$, there are no such 
`canonical' coordinates, since the reduced density matrix cannot be factorized. 
However, one can show that there are indeed $N-n$ eigenstates of the form
\begin{equation}
\psi_{1i} \left( \mathbf{x} \right) = c_{1i} \mathbf{v}_i^T \mathbf{x} \exp \left( - \frac{1}{2} \mathbf{x}^T {\cal W} \mathbf{x} \right) ,
\label{eq:spectrum_first excited_state}
\end{equation}
where the vectors $\mathbf{v}_i$ are the eigenvectors of the matrix
\begin{equation}
\Xi = \tilde{\beta}^T \left( I + {\cal W} \right)^{- 1} .
\label{eq:spectrum_Xi_def}
\end{equation}
Let $\xi_i$ be the eigenvalue of the matrix $\Xi$ 
corresponding to the eigenvector $\mathbf{v}_i$.
Then, the eigenvalue of the eigenstate $\psi_{1i}$ of the reduced density matrix is equal to $\lambda_0 \xi_i$, where $\lambda_0$ is the eigenvalue of the `ground' eigenstate \eqref{eq:eigenstate_ground}. 
Notice that the vectors $\mathbf{v}_i$ are in general complex and the matrix $\Xi$ 
is neither Hermitian or symmetric. Nevertheless, its eigenvalues are real.

Based on the eigenvectors of the matrix $\Xi$, one can build inductively the whole tower of eigenstates of the reduced density matrix. Initially, 
it can be checked that the simple procedure of adding factors $\mathbf{v}_i^T \mathbf{x}$ to the state 
does not result in new eigenstates of the reduced density matrix, i.e. the states
\begin{multline}
\psi_{\left\{ m_1 , m_2 , \ldots , m_n \right\}} \left( \mathbf{x} \right) \\
= c_{\left\{ m_1 , m_2 , \ldots , m_n \right\}} \left( \mathbf{v}_1^T \mathbf{x} \right)^{m_1} \left( \mathbf{v}_2^T \mathbf{x} \right)^{m_2} \ldots \left( \mathbf{v}_{N-n}^T \mathbf{x} \right)^{m_{N-n}} \exp \left( - \frac{1}{2} \mathbf{x}^T {\cal W} \mathbf{x} \right) 
\label{eq:spectrum_general_excited_state}
\end{multline}
are not eigenstates. However, it can be shown that, for each of these states, 
there is a unique way to correct the polynomial 
$\left( \mathbf{v}_1^T \mathbf{x} \right)^{m_1} \left( \mathbf{v}_2^T \mathbf{x} \right)^{m_2} \ldots \left( \mathbf{v}_n^T \mathbf{x} \right)^{m_n}$, adding terms of \emph{smaller} order, so that one gets an eigenstate of the reduced density matrix. 
Then, the corresponding eigenvalue equals
\begin{equation}
\lambda_{\left\{ m_1 , m_2 , \ldots , m_n \right\}} = \lambda_0 \xi_1^{m_1} \xi_2^{m_2} \ldots \xi_{N-n}^{m_{N-n}} .
\label{eq:spectrum_reduced_eigenvalues}
\end{equation}
A more detailed presentation of these results is given in \cite{preparation}.

The above imply that the spectrum of the reduced density matrix is of the form \eqref{eq:reduced_spectrum}, where $\xi_i$ are the eigenvalues of the matrix $\Xi$ given by eq. \eqref{eq:spectrum_Xi_def}.
The entanglement entropy can be expressed as
\begin{equation}
S = - \sum_{i=1}^M \left( \ln \left( 1 - \xi_i \right) + \frac{\xi_i}{1 - \xi_i} \ln \xi_i \right) .
\label{entropyfinal}\end{equation}

\subsection{Entanglement entropy at late times}

We focus now on the limit $\tau\to 0^-$ for a dS background, where we expect
an enhancement of the entanglement entropy, following the paradigm of the two oscillators. 
In this limit, the function $b(\tau,\omega_k)$
can be approximated through eqs. (\ref{bbpappr}), so that eq. (\ref{Omegai}) becomes
\be
{\rm Re}\,\tilde{\Omega}_k\simeq {\cal O}(\tau^2), \quad\quad 
{\rm Im}\,\tilde{\Omega}_k\simeq \frac{1}{\tau} + {\cal O}(\tau).
\label{Omegakiappr}
\ee
To leading order in $\tau$, the matrix $\tilde{\Omega}$ 
of eq. (\ref{Omegatilde}) is the
identity matrix multiplied by $i/\tau$. As a result, the leading contribution to the 
matrix $\Omega = O^T \tilde{\Omega}\, O$ of eq. (\ref{Omega}) is the same as for
$\tilde{\Omega}$. This implies that
\be
{\rm Re}\,\Omega\simeq {\cal O}(\tau^2), \quad\quad 
{\rm Im}\,\Omega\simeq \frac{1}{\tau} I + {\cal O}(\tau).
\label{Omegaiappr}
\ee
The leading contribution to the imaginary part of $\Omega$ survives only 
in the blocks $A$ and $C$ of the matrix $\Omega$, i.e.
\begin{align}
{\rm Re}\,A &\simeq  {\cal O} \left( \tau^2 \right) , \quad\quad 
{\rm Im}\,A \simeq \frac{1}{\tau} I + {\cal O} \left( \tau \right) , \\
{\rm Re}\,B &\simeq  {\cal O} \left( \tau^2 \right) , \quad\quad 
{\rm Im}\,B \simeq  {\cal O} \left( \tau \right) , \\
{\rm Re}\,C &\simeq  {\cal O} \left( \tau^2 \right) , \quad\quad 
{\rm Im}\, C \simeq \frac{1}{\tau} I + {\cal O} \left( \tau \right) .
\end{align}
The definitions of the matrices $\gamma$, $\delta$ and $\beta$ (\ref{gammam}) and (\ref{betam}) then directly imply that at late times
\begin{align}
\gamma &\simeq {\cal O} \left( \tau^0 \right) , \\
\delta &\simeq \frac{1}{\tau} I + {\cal O} \left( \tau \right) , \\
\beta &\simeq {\cal O} \left( \tau^0 \right)
\end{align}
and that at leading order
\begin{equation}
\beta \simeq \gamma \simeq \frac{1}{2} {\rm Im}B^\dagger \,
{\rm Re} \left( A \right)^{-1}\, {\rm Im}B . \label{leading}
\end{equation}
The leading contribution 
appears only in the matrix $\delta$, which 
does not affect the eigenvalues of the reduced density matrix. 
These conclusions are consistent with the explicit expressions (\ref{densres})-(\ref{deltar}) in the case of two oscillators. 

The fact that the matrices $\beta$ and $\gamma$ become equal as $\tau \to 0^-$ would suggest that the matrix $\tilde{\beta}$ tends to the identity matrix. 
It follows that the matrix ${\cal W}$, given by eq. \eqref{eq:omega}, would tend to a vanishing matrix, 
and the matrix $\Xi$ of eq. \eqref{eq:spectrum_Xi_def} to the identity. 
This would imply that all the $N-n$ eigenvalues $\xi_i$ tend to 1, 
generating a term to the entanglement entropy
which is proportional 
to the number  $\left( N - n \right)$ of the degrees of freedom of the subsystem, 
i.e. it is a \emph{volume} term.

However, this expectation is too naive to hold in all cases. 
The reason is that the above argument is valid 
only if the matrix $\gamma$ is invertible. 
Only then the matrix $\tilde{\beta}$ tends to the identity matrix. 
The time evolution of the wave functions of the normal modes in a 
dS background provides an example in which this expectation is not realized. 
The crucial point is that the expansion of the complex quantities defined in eq. (\ref{Omegai})
for late times is very specific, namely
\begin{equation}
{\rm Im}\,\tilde{\Omega}_k\simeq \frac{1}{\tau} - \omega_k^2 \tau .
 \end{equation}
This implies that the expansion of the 
imaginary part of the matrix $\Omega$ defined in eq. (\ref{Omega})
is 
\begin{equation}
{\rm Im}\,{\Omega} \simeq  \frac{1}{\tau} I - \tau K ,
\label{matrixK} \end{equation}
where $K$ is the matrix of couplings that appear in the original discretized Hamiltonian, 
such as, for example, the Hamiltonian (\ref{eq:Hamiltonian_discretized}).
The off-diagonal block matrix $B$ in the notation of eq. (\ref{block-form}), which connects oscillators of subsystem 1 to oscillators of subsystem 2, tends to a matrix that contains non-vanishing elements at order $\tau$ only corresponding to the pairs of oscillators
that are directly coupled via a Hamiltonian term. 
For the Hamiltonian (\ref{eq:Hamiltonian_discretized}), 
which emerges from a local field theory,
such elements are non-vanishing only between degrees of freedom at
neighbouring sites. 

It follows that
the matrix $\gamma$ does not always tend to an invertible matrix, and only as many $\xi_i$ as the number of non-vanishing elements of the block $B$ at order $\tau$ tend to 1. 
As a result, a volume term would be possible only in a non-local field theory, which would contain direct couplings between non-neighbouring oscillators. In a local field theory, the divergent term is necessarily proportional to the area of the entangling surface and not 
to the volume of the considered subsystem.

However, a more general behaviour is possible. The conclusion we reached for
a dS background 
is due to the specific form of the function 
$\tilde{\Omega}_k \left( \tau \right)$ in this case. 
In the case of a RD or MD background the form of $\tilde{\Omega}_k \left( \tau \right)$ is
drastically different. 
It is, therefore, possible that volume terms may emerge during these periods.
We shall find indications for such an effect in the following section.

Defining $\xi_i = 1 - \epsilon_i$ for the $\xi_i$ which tend to one at late times, it is easy to show that
\begin{equation}
S = \sum_i \left( - \ln \epsilon_i + 1 + {\cal O} \left( \epsilon_i \right) \right) .
\label{eq:SEE_expansion}
\end{equation}
As an indicative example, in the case of the two oscillators we have
\begin{equation}
\epsilon(\tau) =  \left|\frac{4\, (\omega_+\omega_-)^{3/2}}{\omega^2_- - \omega_+^2} \tau \right|.
\label{eq:epsilon}
\end{equation}
In general, the expression \eqref{eq:SEE_expansion} implies that the entanglement entropy at late times diverges as
\begin{equation}
S \sim \ln \tau \sim {\cal N},
\end{equation}
where ${\cal N}=\ln a(\tau)= -\ln|H\tau|$ is the number of efoldings.

As a final comment, we would like to mention a subtlety of the field-theoretical setup
that becomes important at this point. 
In our discussion up till now we have assumed that the total number of 
degrees of freedom remains constant. The theory of the quantum field includes
a UV cutoff $\epsilon$ that regulates divergent contributions to physical quantities. 
The number of degrees of freedom increases for decreasing $\epsilon$. This is 
apparent if $\epsilon$ is identified with the lattice spacing in a
discretization of the field. In this respect, it becomes crucial how one defines
the UV cutoff of the theory. For a quantum field on an expanding background, 
the total number of degrees of freedom would remain constant if the UV cutoff
was defined in comoving coordinates, which would then imply that the physical 
length of the cutoff grows with the scale factor. However, in the more intuitive approach 
that the UV cutoff is identified
with a physical scale, such as the Planck scale, the UV sector of the theory is
continuously replenished with new modes. In the second picture, the 
entanglement entropy would still display a contribution $\sim  1/\epsilon^2$, exactly
as in the static case. Our analysis reveals the presence of a new divergent term,
arising through the modes that get stretched beyond the horizon. 
The number of such modes grows with the duration of inflation and the resulting
divergence for ${\cal N} \to \infty$ is an IR one.

\section{Entanglement entropy of a quantum field in $1+1$ dimensions} \label{2d}

\subsection{The model}

The calculation of the entanglement entropy of a quantum field in $3+1$ dimensions 
is hindered by several technical difficulties, whose resolution requires a separate study.
For example, the regularization of the contributions from UV angular modes, corresponding
to large values of $l$ and $m$ in eq. (\ref{eq:Hamiltonian_discretized}), must be done carefully, so as
to preserve the statistical homogeneity of the problem. This is important in order to 
identify correctly possible volume effects. Moreover, the efficiency of the numerical
implementation must be maximized for the calculation to be feasible. 
As we discussed at the end of the previous section, there are also 
more fundamental issues concerning the number, as well as the state, of the 
UV and IR modes in an expanding universe that
possesses a {\it physical} UV cutoff, such as the Planck scale. 

It is possible, however, to carry out a complete numerical calculation in a toy model of 
a massless scalar field in $1+1$ dimensions in order to confirm the various 
features of the entanglement entropy expected from the discussion in the previous
sections. In $1+1$ dimensions, if one assumes a background given by the FRW metric of the form (\ref{dsmetric}),
neglecting the angular part, the scalar field is canonically normalized. The inclusion of
an effective mass term that depends on the background 
can be achieved by allowing for a non-minimal
coupling to gravity of the form $\xi R \phi^2$. For a dS background with $a(\tau)=-1/(H\tau)$,
the curvature scalar $R$ in $1+1$ dimensions is equal to $-2H^2$. 
The choice $\xi=-1/2$ results in an action for the field $\phi(\tau,\bx)$ of the form (\ref{action2}) with $\kappa=1$. 
We also saw in section \ref{QCS} that during the RD era we have $\kappa=0$. A similar behaviour 
can be introduced in our $(1+1)$-dimensional setup by assuming a transition to a
flat background with $R=0$ at some time $\tau_0$. 

These considerations lead to a discretized system described by the Hamiltonian
of a chain of oscillators with next-neighbour couplings and time-dependent frequencies:
\be
H = \frac{1}{2\epsilon} 
\sum\limits_{j = 2}^{N-1} \left[ \pi _{j}^2 +
\left( f_{j + 1} - f_{j} \right)^2 
-\frac{2\kappa}{(\tau/\epsilon)^2} f_{j}^2 \right] 
+\frac{1}{2\epsilon}
\sum\limits_{j = 1,N} \left[ \pi _{j}^2 +
 f_{j}^2  
-\frac{2\kappa}{(\tau/\epsilon)^2} f_{j}^2 \right] .
\label{eq:Hamiltonian_discretized2}
\ee
We have modified the action for the oscillators at the ends of the chain, so as to
impose boundary conditions corresponding to a vanishing field at the endpoints.
In the picture of coupled oscillators, the ones at the ends of the chain are
connected to fixed surfaces.
Other types of boundary conditions are also possible, leading to similar results for
the entanglement entropy.

The frequency-squared matrix, whose diagonalization defines the canonical modes, is 
of the tridiagonal form
\be
\left(
\begin{array}{ccccc}
2 & -1 & 0 & 0 & \\
-1 & 2 & -1 & 0 & \\
0 & -1 & 2 & -1 & \cdots \\
0 & 0 & -1 & 2 &  \\
 &  & \vdots &  & \\
\end{array} \right).
\label{triagonal}
\ee
In our normalization all dimensionful quantities are expressed in units
of the comoving lattice spacing $\epsilon$, with $1/\epsilon$ acting as a UV cutoff.
Thus, our normalization corresponds to setting $\epsilon=1$. 
The eigenfrequencies and eigenfunctions of the above matrix are well known. 
The frequencies are given by 
\be
\omega_k=2 \sin \left(\frac{k \pi}{2(N+1)} \right), 
~~~~~~~~~
k=1,2,\cdots, N.
\label{eigenfr2} \ee
For large $N$, they vary between ${\cal O}(1)$ and ${\cal O}(1/N)$ 
in units of $\epsilon$, with the wavelengths having the inverse 
variation between ${\cal O}(1)$ and ${\cal O}(N)$. 

Tracing out part of the chain in order to derive the reduced density matrix
is not possible through analytical means. For this reason, we rely on 
a numerical study of this system. We consider a dS era with $\kappa=1$ in eq. 
(\ref{eq:Hamiltonian_discretized2}), followed by an era with $\kappa=0$ that mimics the
RD era in $3+1$ dimensions. Notice that for $\tau\to -\infty$ we recover the Hamiltonian 
of a free massless scalar in a static background. 
The system corresponds to the discretized version of a
conformal field theory with central charge $c=1$, for which the entanglement entropy is
explicitly known \cite{wilczek,korepin,cardy1,cardy2}.
For a finite system of physical 
length $L$ with boundaries, divided into two pieces of lengths $l$ and $L-l$, the 
entanglement entropy is \cite{cardy2}
\be
S=\frac{c}{6}\ln\left(\frac{2L}{\pi \epsilon} \sin \frac{\pi l}{L} \right) +\bar{c}'_1, 
\label{entropycardy} \ee
with $\bar{c}'_1$ being a scheme-dependent constant.

\begin{figure}[t!]
\centering
\includegraphics[width=0.8\textwidth]{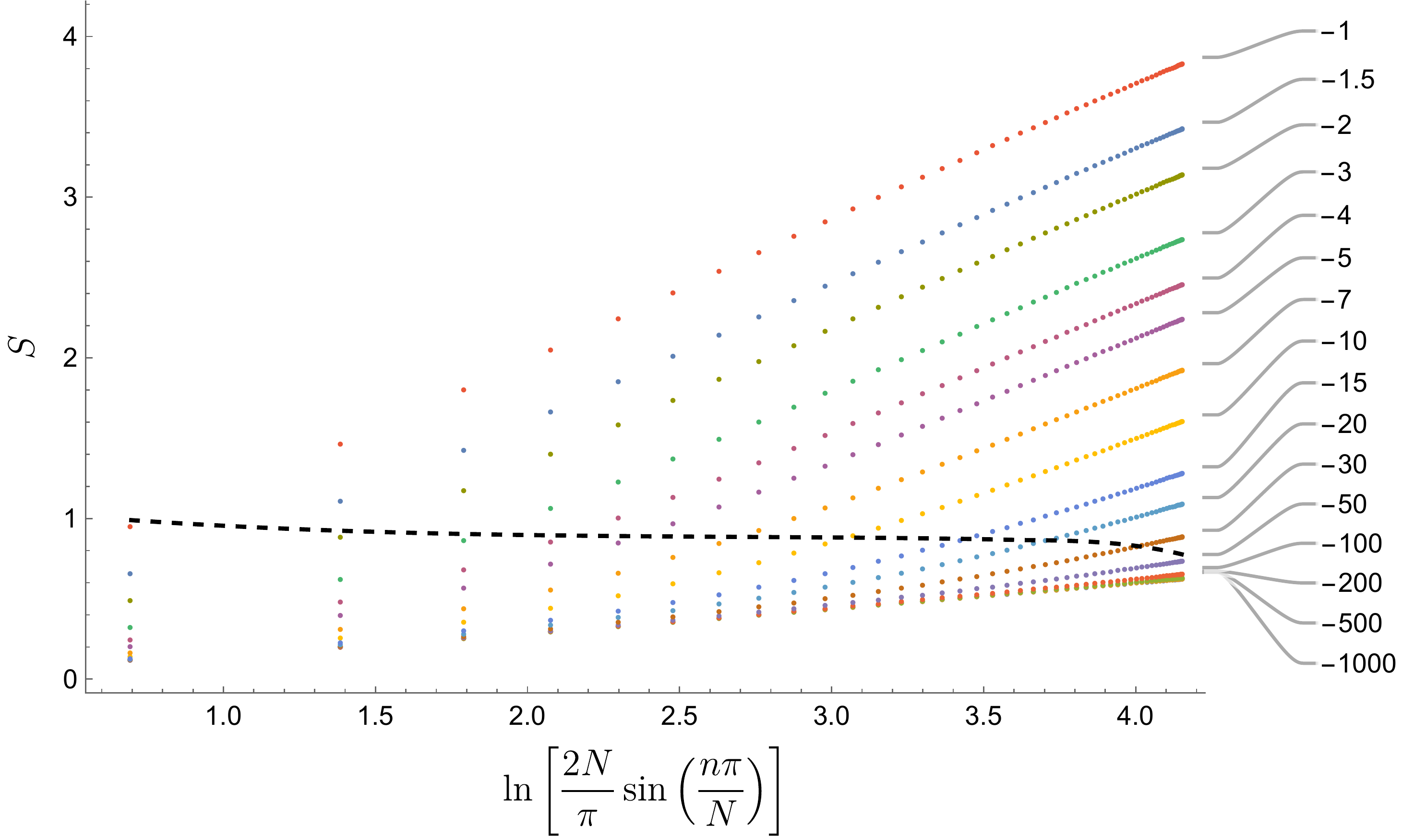} 
\caption{
The entanglement entropy resulting from tracing out the part $n<k\leq N$  
of a one-dimensional chain at
various times, for a dS background. 
}
\label{2ddS}
\end{figure}

For the calculation of the entanglement entropy we follow the steps outlined in the
previous sections. We first compute the eigenfrequencies $\omega_k$ 
and eigenfunctions of the 
matrix (\ref{triagonal}), from which we deduce the complex quantities (\ref{Omegai}). 
The function $b(\tau,\omega_k)$ for a system in its ground state 
is given by eq. (\ref{b1}) for exact dS space, and by
eqs. (\ref{bdS2sf}), (\ref{brd}) for a transition from a dS to a RD regime. We then
switch from the canonical variables corresponding to the eigenfunctions of (\ref{triagonal}) 
to the basis of the degrees of freedom at each point on the chain, thus
obtaining the matrix $\Omega$ of eq. (\ref{Omega}). Tracing out part of the chain
leads to the matrices $\gamma$ and $\beta$ of eqs. (\ref{gammam}), (\ref{betam}), from
which we compute the matrix $\tilde{\beta}$ of eq. (\ref{betatildem}). We then
find the eigenvalues of the matrix $\Xi$ defined through eqs. (\ref{eq:omega}), (\ref{eq:spectrum_Xi_def}), carefully selecting the matrix ${\cal W}$ that corresponds to the
normalizable state (\ref{eq:eigenstate_ground}). Finally, we compute the 
entanglement entropy through eq. (\ref{entropyfinal}).

\subsection{Entanglement entropy during inflation}

The results for a dS background are presented in figure \ref{2ddS} at various conformal 
times $\tau$
ranging from $-1000$ to $-1$. We depict the entropy as a function of the part of the
chain that is kept after tracing out the remainder, 
in terms of the quantity $s=\ln[(2N/\pi)\sin(n \pi/N)]$, 
with $n=1,2,\cdots,N$, so as to
be able to make a direct comparison with eq. (\ref{entropycardy}).
Using this variable results in the depiction of the entropy from parts of the chain up 
to its middle. 
Longer parts result in values of the entropy equal to those of their complementary parts,
as required by the symmetry property of entanglement entropy for 
pure states of the overall system.
The equality of the entropies for the two pieces with lengths $l$ and $L-l$
of a chain with total length $L$ is confirmed by our calculation with high accuracy. 

We observe that at very early times the entropy is a linear function of $s$. The slope
can be determined through a numerical fit. It lies within a few-percent range from 
the expected value $1/6$, with agreement that improves with increasing $N$. In figure
\ref{2ddS} we depict the results for $N=100$, but we have computed the entropy for
larger values of $N$ as well. 
We emphasize that the various lengths correspond to comoving distances. The physical
lengths increase proportionally to the scale factor. However, this also happens for
the lattice spacing, resulting in a UV cutoff that is reduced with increasing time.  
Both ratios $L/\epsilon$ and $l/L$ are the same for comoving and physical lengths, which
explains the agreement with eq. (\ref{entropycardy}).

At later times we observe a deviation from a linear relation, which starts from 
large values of $n$. In order to understand this behaviour, we have included in 
figure \ref{2ddS} a dashed line that indicates the location of the horizon length $1/H$,
defined through the relation $n\,\epsilon\, a(\tau)/(1/H)=1$. For $a(\tau)=-1/(H \tau)$,
this gives $n=-\tau\epsilon$. For smaller $|\tau|$, increasingly larger parts of the
chain extend beyond the horizon. More and more canonical modes of the system 
becomes squeezed, leading to the increase of the entanglement entropy.
For $\tau\to 0^-$ the function $S(s)$ becomes linear again, with a slope that approaches
1. Numerical fits indicate a few-percent agreement with this value, improving for
larger $N$. The interpretation of this result in the context of a continuous field theory
is less clear. As can be seen in figure \ref{2ddS} the cutoff length $1/\epsilon$ becomes
comparable to the horizon length. The discretization of the system is sparse and the
effect of the UV modes with very small {\it physical} wavelength is lost. This
is a result of our inability to reproduce the limit $N\to \infty$, or $\epsilon \to 0$, numerically.

However, the overall increase of the entropy at late times, expected from our discussion
in the previous section, is clearly seen in figure \ref{2ddS}. The numerical solution 
verifies with very good accuracy the linear increase with $\ln(|\tau|)$. 
More specifically, the entanglement entropy for $\tau \to 0^-$ can be described very well by a 
function 
\be
S=\ln\left(\frac{2L\,a(\tau)}{\pi \epsilon} \sin \frac{\pi l}{L} \right) +d, 
\label{entropycardytime} \ee
with constant $d$, where we have made use of the relation $a(\tau)=-1/(H\tau)$.
This expression can be interpreted as representing the contribution to the entanglement 
entropy arising from the deep IR, where squeezing plays a dominant role.
We observe that a volume term, namely a term proportional to the length of the considered subsystem, 
does not emerge during this period, as expected from our arguments in the previous section.

Even though it is not possible to reproduce numerically the limit $N\to \infty$ or 
$\epsilon\to 0$, 
figure \ref{2ddS} clearly demonstrates the presence of two regimes in the form 
of the entanglement entropy. For regions much smaller than the horizon, the entanglement
entropy is dominated by the UV contributions, which result in the standard 
result (\ref{entropycardy}) for a static background. For regions comparable 
to the horizon the IR modes become important and the entropy is time-dependent.
When the IR modes dominate, as for regions extending beyond the horizon,
the extrapolation of our results points towards a form of the entropy described by
the formula (\ref{entropycardytime}).

\begin{figure}[t!]
\centering
\includegraphics[width=0.6\textwidth]{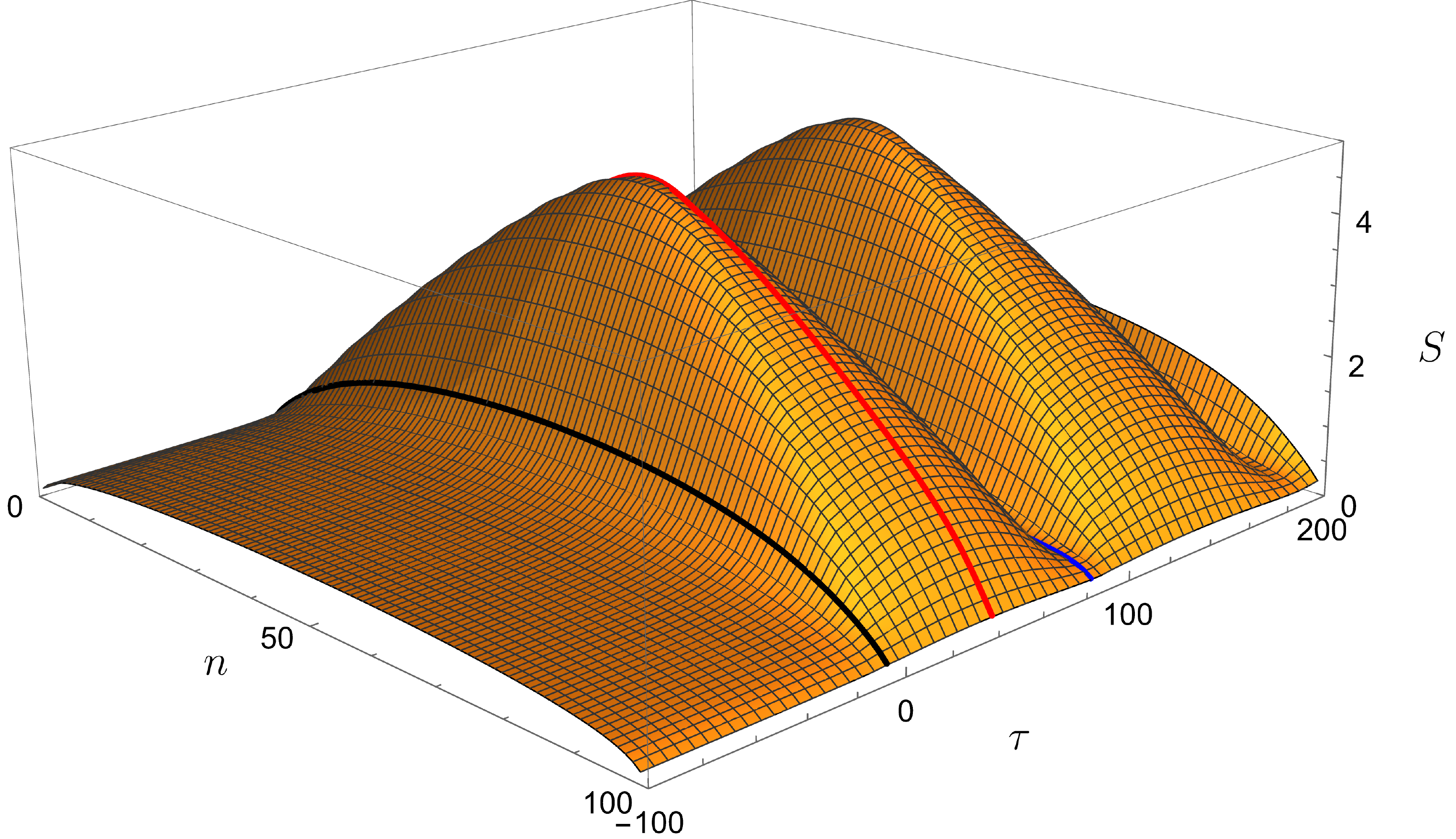} 
\includegraphics[width=0.38\textwidth]{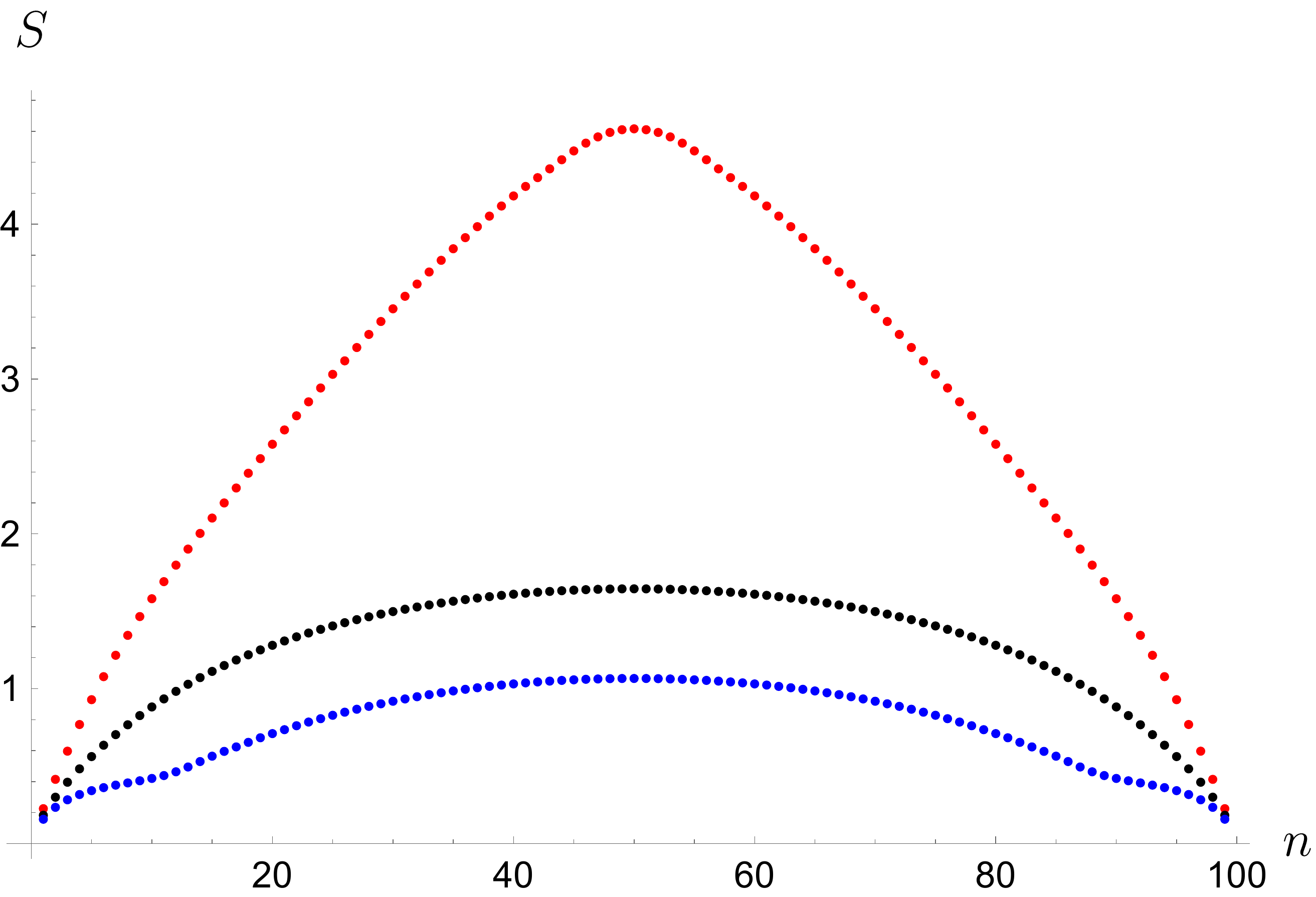} 
\caption{
The entanglement entropy resulting from tracing out the part $n<k\leq N$ 
of a one-dimensional chain at
various times. The transition from a dS to a RD background (black line) occurs at  
$\tau_0=-5$. During the RD era, the maximal entanglement entropy (red line) 
is first achieved at $\tau=85$, and
the minimal entanglement entropy (blue line) at $\tau=40$. For clarity, we also display
the entanglement entropy at these times in the right plot.
}
\label{2ddSRD}
\end{figure}

\subsection{Entanglement entropy during radiation domination}

The next question of interest concerns the evolution of the entanglement entropy during
a RD era that follows inflation. Our toy model can provide intuition on this issue,
even though the RD regime is only mimicked by our construction.
In figure \ref{2ddSRD} we depict the entanglement entropy as a function of the 
part of the chain that has not been traced out and the conformal time. We have used
$N=100$, $\epsilon=1$, $H=0.1$ during the dS era, and $\tau_0=-5$ for the time of 
transition to the RD era. We have selected values such that the bigger part of the 
chain lies outside the horizon at $\tau_0$, while there is sufficient
discretization of the subhorizon part. 
As we plot the entanglement entropy as a function of $n$, the whole chain is
visible. 
The entanglement entropy at the transition from a dS to a RD background at  
$\tau_0=-5$ is depicted by the black line. 

At early times the plot reproduces the function (\ref{entropycardy}), as we explained above. 
At later times during the dS era, the entanglement entropy increases. After the transition 
to the RD era, a further increase is followed by an oscillatory pattern. 
The maximal entanglement entropy (red line) 
is first achieved at $\tau=85$, and
the minimal entanglement entropy (blue line) at $\tau=40$. For clarity, we also display
the entanglement entropy at these times in the right plot. The behaviour
is very similar to that depicted in figure \ref{fig8} for the case of two coupled 
oscillators. In the current example, there exist $N=100$ canonical modes, parts of which
are traced out. It is very difficult to predict the effect of the various eigenfrequencies
on the final oscillatory pattern in the entanglement entropy. The numerical calculation 
indicates that the pattern is dominated by the lowest-frequency mode with 
a wavelength of ${\cal O}(N)$. This is not surprising, as this the mode that `exited the
horizon' first during the dS era, and thus its squeezing has been maximally enhanced.

The general conclusion that can be reached is that, after the end of the 
inflationary era, the canonical modes of the overall system cannot 
return to minimal uncertainty states. The modes remain in squeezed states, with
wave functions that oscillate 
periodically. Thus, the entanglement due to these modes fluctuates without 
having a decreasing trend; on average it remains high.

An interesting quantitative feature of figure \ref{2ddSRD} concerns the form of the 
dependence of the entropy on the length of the part of the chain that is not
traced out. During the inflationary era this dependence is purely logarithmic and
does not indicate any significant volume contributions to the entropy. 
However, in the RD era and during the periods of large squeezing, the dependence becomes
much sharper. At the times of maximal squeezing and entropy, the dependence is almost linear.
In 1+1 dimensions, this feature results from contributions proportional to the volume, 
which involve the total number of degrees of freedom of the subsystem. In the
previous section we speculated that such an effect is possible during the RD era.

\section{Summary and conclusions} \label{concl}

In this work we investigated the effect of the expansion of spacetime on
the entanglement entropy associated with quantum fields. Our aim was to understand 
whether a significant imprint can be left on a quantity that is sensitive to 
the quantum nature of the fields. We were especially interested in the period of inflation, 
during which the expansion is very rapid and the effects can be enhanced. 
The standard picture of the evolution of the field modes during inflation 
points in the opposite direction. Momentum modes that start as pure quantum 
fluctuations in the Bunch-Davies vacuum are expected to freeze when they exit the
horizon and transmute into classical stochastic fluctuations. 
We have seen that this is only part of the picture. Even though its classical features are dominant, the field never loses its quantum nature. Moreover, 
it is well known that the various modes evolve into squeezed states. 
Our main result is that the squeezing triggers a strong enhancement of quantum entanglement.
The effect is clearly visible in the entanglement entropy. 
 
The entanglement entropy is obtained by dividing the 
degrees of freedom of a system into two subsets and tracing out one of them
in order to extract the reduced density matrix for the other.
We analysed here the case that these degrees of freedom are local and correspond
to the interior or exterior of an entangling surface. In flat space, 
the entanglement entropy for a scalar field is dominated by an area law term,
similarly to the black hole entropy. We investigated how this result is affected
by the expansion of the background spacetime. The analysis requires an extension of
the techniques developed by Srednicki in his seminal study of the entanglement 
entropy \cite{srednicki}. 

A first necessary step is the determination of the 
wave functions that describe the quantum harmonic oscillators corresponding to the
canonical modes of the field in an expanding background. This task can be performed
in a straightforward manner for a dS background, implementing appropriate initial
conditions corresponding to the Bunch-Davies vacuum.  
The system is then assumed to lie in the ground state of 
each canonical mode in the asymptotic past. The calculation can also be extended to 
a subsequent RD or MD phase. Tracing out the local modes inside or outside an entangling
surface is again straightforward, even though it must be performed numerically. 
The most challenging part is the determination of the eigenvalues and eigenstates
of the reduced density matrix, which is much more involved than in the static case. 
This task required a separate extensive analysis, which
is presented in \cite{preparation}. In section \ref{QF} we summarized
the main results, which provide the basis for the numerical calculation. 
We also presented the emerging qualitative features of the entanglement entropy.

We performed two explicit calculations of the entanglement entropy. The first one was
in the context of a toy model of two harmonic oscillators, with 
frequencies that have the same time dependence as the modes of a scalar field in
an expanding background. The tracing out of one of the oscillators in order to 
derive the reduced density matrix of the other and the entanglement entropy can be
done analytically in this case. The second calculation was done in the context of 
a (1+1)-dimensional field theory, with an action that displays the same time dependence
as for a field in a (3+1)-dimensional dS or RD background. Even though the calculation
could only be performed numerically, its validity was checked through the comparison
with known analytical results for the entanglement entropy of conformal field theories
in (1+1)-dimensional flat space. Because of its higher technical complexity,  
the numerical calculation of the entropy in a 
(3+1)-dimensional field theory is deferred to future work.

Our main result is that the entanglement entropy during the inflationary era is an 
increasing function of time. More specifically, it increases proportionally to 
the number of efoldings, as we discussed at the end of section \ref{QF}. 
We emphasize that the effect is not limited to the 
number of efoldings that can be probed through classical observables in the 
present-day universe, such as the CMB spectrum. The entanglement entropy 
is actually affected by the total duration of inflation. Thus it could provide 
a means for exctracting information for 
periods of inflation that are {\it inaccessible to other probes}.
 
The increase of entanglement is a consequence of the canonical modes of the total 
system evolving to squeezed states when their characteristic physical wavelength becomes
larger than the horizon radius. Thus, the well known freezing of these modes is 
intrinsically linked to the squeezing of their quantum states and their increased 
entanglement.
A second important conclusion from our analysis is that, after the end of the 
inflationary era and during an era of radiation or matter domination, 
the wave functions of the canonical modes do not 
retreat to minimal uncertainty states. The modes remain in squeezed states that oscillate 
periodically.  
The entanglement due to these modes fluctuates, but does not have a decreasing trend. 
On average it remains high. Therefore, there is a quantum effect of the 
inflationary era, proportional to the number of modes that became frozen at some time, independently of whether they became unfrozen during a later stage.
This effect cannot be described through classical 
stochastic fluctuations, while it is imprinted on the entanglement entropy 
in the post-inflationary universe.

We point out that a distinction must be made between the eigenfrequencies of the
overall system and those of the reduced one.
The reduced system contains fewer  
degrees of freedom, which
also compose effective canonical modes. 
Each of them has an effective eigenfrequency, which is 
a function of the eigenfrequencies of the canonical modes of the overall system. 
The effective eigenfrequencies set a critical time after which 
the contribution of the effective mode 
to entanglement becomes large. The relation between 
the eigenfrequencies of the overall system and the those of the reduced one is
complicated. For example, in the case of the two oscillators, 
the reduced system has a single degree of freedom and a 
single effective mode, which starts enhancing the 
entanglement entropy at the 
critical time given by (see eq. \eqref{eq:epsilon})
\begin{equation}
\tau_{} =- \frac{|\omega^2_- - \omega_+^2|}{4\, (\omega_+\omega_-)^{3/2}} .
\end{equation}
This critical time does not coincide with the time 
the modes of the overall system cross the horizon, 
but it is a function of them. For a discretized field theory with a large 
number of degrees of freedom, the corresponding relations are very complicated and
can be determined only numerically. However, one would expect that it is the 
lowest frequency modes of the overall system that give the largest effect, as they
are the first ones to become superhorizon and, therefore, maximally squeezed.
For example, in the above relation it is the mode of lowest frequency that 
dominates the entanglement entropy. A similar effect can be seen in figure \ref{2ddSRD}
at late times during the RD era. The oscillatory form of the entropy has a characteristic
frequency set by the lowest frequency of the overall system, which is set by the total length
of the chain in this case. 

The above considerations reveal some conceptual issues that must be 
addressed in the study of the entanglement entropy of our universe. 
If the volume of the overall system is unbounded, there is no lower limit for 
the eigenfrequencies, resulting from the size of the system. In such a case, the 
lowest-frequency mode that becomes maximally squeezed, and thus dominates 
the entropy, is the one that exited the horizon at the
beginning of inflation. It is an exciting prospect that the entanglement entropy may
encode information for an event that took place much earlier than the stages of 
inflation probed
by other means. However, extracting this information is not an easy task.
In this respect, the dependence of the entropy on the size of the entangled region 
can be important. As was observed in figure \ref{2ddSRD}, a volume contribution appears in 
the RD era,
which is maximal at the times
when the entropy becomes maximal. This effect may persist in 3+1 dimensions, providing
significant modifications of the area law, and thus a signature of the inflationary
expansion.

Apart from the above IR issue, the effect of the UV modes on 
the entanglement entropy becomes complicated during inflation. 
We bypassed this issue by keeping a fixed {\it comoving} UV cutoff, equal to
the comoving lattice spacing, in the discretized field theory.  
However, the standard assumption is that the expanding universe 
possesses a {\it physical} UV cutoff, such as the Planck scale. 
This requires that the comoving lattice spacing shrink with time.  
As a result, the number of degrees of
freedom in a comoving volume increases, as new modes emerge from the Planck scale.
It is not obvious how to define the quantum state of the system in terms of the
canonical modes, as the canonical variables must be modified continuously during
the evolution in order to incorporate the new degrees of freedom. 
A possible resolution of this issue is that the new UV modes simply add new canonical
modes of very high frequency, for which the expansion is irrelevant. As a result, 
the entanglement entropy will still include the standard divergent contribution,
regulated by the {\it physical} UV cutoff. The novel effect of the expansion that we
analyzed here arises through the squeezing of the low-frequency modes with 
wavelengths comparable to the horizon. An implementation of these ideas and the
calculation of the entanglement entropy in an expanding (3+1)-dimensional universe
is the subject of ongoing work.

\appendix
\section{The influence of the mass of the field during the inflationary period} \label{appendixA}

In the main body of this work, we focussed on the case of a massless scalar field. In this case, the function $b \left( \tau \right)$, which determines the evolution of the wave function during the inflationary period, assumes a simple form. In the more general case of a massive field, the function $b \left( \tau \right)$ is given by eq. \eqref{bsolfin} in terms of the Bessel functions of the first and second kind. For real values of its index, the Bessel function of the first kind has a series representation of the form
\begin{equation}
J_\nu \left( x \right) = \sum_{n = 0}^\infty c_n x^{\nu + 2 n} .
\end{equation}
This does not hold for the series representation of the Bessel function of the second kind. This has a double series representation of the form
\begin{equation}
Y_\nu \left( x \right) = \sum_{n = 0}^\infty c_n x^{\nu + 2 n} + c_n^\prime x^{- \nu + 2 n} .
\end{equation}
It directly follows from eq. \eqref{bsolfin} that the function $b \left( \tau \right)$ has in general a triple series representation of the form
\begin{equation}
b \left( \tau \right) = \sum_{n = 0}^\infty c_n \tau^{2 \nu + 1 + 2 n} + c_n^\prime \tau^{1 + 2 n} + c_n^{\prime\prime} \tau^{- 2 \nu + 1 + 2 n} .
\end{equation}
The function $b \left( \tau \right)$ can be analytically continued for imaginary indices. It is a matter of tedious algebra to show that in general
\begin{equation}
b^2 \left( \tau \right) = \begin{cases}
\frac{1}{\omega_0^2 \tau^2} + 1 , & \kappa = 1 \\
\frac{\Gamma^2 \left( \nu \right)}{\pi} \left( - \frac{\omega_0 \tau}{2} \right)^{1 - 2 \nu} + \frac{2 \Gamma^2 \left( \nu \right)}{\left( \nu - 1 \right) \pi} \left( - \frac{\omega_0 \tau}{2} \right)^{3 - 2 \nu} + \mathcal{O} \left( \tau \right) , & \frac{3}{8} < \kappa < 1 , \\
\frac{\Gamma^2 \left( \nu \right)}{\pi} \left( - \frac{\omega_0 \tau}{2} \right)^{1 - 2 \nu} - \frac{\cot \left( \nu \pi \right)}{\nu} \left( - \frac{\omega_0 \tau}{2} \right) + \mathcal{O} \left( \tau^{3 - 2 \nu} \right) , & 0 < \kappa < \frac{3}{8} , \\
1 , & \kappa = 0 , \\
\frac{\Gamma^2 \left( \nu \right)}{\pi} \left( - \frac{\omega_0 \tau}{2} \right)^{1 - 2 \nu} - \frac{\cot \left( \nu \pi \right)}{\nu} \left( - \frac{\omega_0 \tau}{2} \right) + \mathcal{O} \left( \tau^{1 + 2 \nu} \right) , & - \frac{1}{8} < \kappa < 0 , \\
\frac{ \left| \Gamma \left( i | \nu | \right) \right|^2}{\pi}
\left( - \omega_0 \tau \right) \big[ \cosh \left( \pi | \nu | \right) & \\
+ \cos \left( 2 | \nu | \ln \frac{-\omega_0 \tau}{2} - 2 {\rm arg} \Gamma\left(i | \nu | \right) \right) \big] + \mathcal{O} \left( \tau^2 \right) , & \kappa < - \frac{1}{8} .
\end{cases}
\end{equation}
The index $\nu$ is given in terms of $\kappa$ via eq. \eqref{nu}.

The above expansion of the function $b \left( \tau \right)$ directly implies the following for the coefficient $\Omega_k$ of the quadratic term in the exponent of the mode wave functions, which is given by eq. \eqref{Omegai}:
\begin{equation}
\mathrm{Re}\, \tilde{\Omega}_k \left( \tau \right) = \begin{cases}
\omega_k^3 \tau^2 - \omega_k^5 \tau^4 , & \kappa = 1 \\
\frac{\pi \omega_k}{\Gamma^2 \left( \nu \right)} \left( - \frac{\omega_k \tau}{2} \right)^{2 \nu - 1} + \frac{2 \pi \omega_k}{\left( \nu - 1 \right) \Gamma^2 \left( \nu \right)} \left( - \frac{\omega_k \tau}{2} \right)^{2 \nu + 1} + \mathcal{O} \left( \tau^{4 \nu - 1} \right) , & \frac{3}{8} < \kappa < 1 , \\
\frac{\pi \omega_k}{\Gamma^2 \left( \nu \right)} \left( - \frac{\omega_k \tau}{2} \right)^{2 \nu - 1} - \frac{2 \pi^2 \cot \left( \nu \pi \right) \omega_k}{\nu \Gamma^4 \left( \nu \right)} \left( - \frac{\omega_k \tau}{2} \right)^{4 \nu - 1} + \mathcal{O} \left( \tau^{2 \nu + 1} \right) , & 0 < \kappa < \frac{3}{8} , \\
\omega_k , & \kappa = 0 , \\
\frac{\pi \omega_k}{\Gamma^2 \left( \nu \right)} \left( - \frac{\omega_k \tau}{2} \right)^{2 \nu - 1} - \frac{2 \pi^2 \cot \left( \nu \pi \right) \omega_k}{\nu \Gamma^4 \left( \nu \right)} \left( - \frac{\omega_k \tau}{2} \right)^{4 \nu - 1} + \mathcal{O} \left( \tau^{- 1 - 2 \nu} \right) , & - \frac{1}{8} < \kappa < 0 , \\
- \frac{\pi}{\left|\Gamma\left(i | \nu | \right)\right|^2 \tau} \frac{1}{\cosh \left( \pi | \nu | \right) 
+ \cos \left( 2 | \nu | \ln \frac{-\omega_0 \tau}{2} - \varphi_\nu \right)} + \mathcal{O} \left( \tau^0 \right) , & \kappa < - \frac{1}{8} 
\end{cases}
\end{equation}
and
\begin{equation}
\mathrm{Im}\, \tilde{\Omega}_k \left( \tau \right) = \begin{cases}
\frac{1}{\tau} - \omega_k^2 \tau , & \kappa = 1 \\
\frac{\left( 2 \nu - 1 \right)}{2 \tau} - \frac{\omega_k^2 \tau}{2 \left( \nu - 1 \right)} + \mathcal{O} \left( \tau^{2 \nu - 1} \right) , & \frac{3}{8} < \kappa < 1 , \\
\frac{\left( 2 \nu - 1 \right)}{2 \tau} - \frac{\pi \cot \left( \nu \pi \right) \omega_k}{\Gamma^2 \left( \nu \right)} \left( - \frac{\omega_k \tau}{2} \right)^{2 \nu - 1} + \mathcal{O} \left( \tau \right) , & 0 < \kappa < \frac{3}{8} , \\
0 , & \kappa = 0 , \\
\frac{\left( 2 \nu - 1 \right)}{2 \tau} - \frac{\pi \cot \left( \nu \pi \right) \omega_k}{\Gamma^2 \left( \nu \right)} \left( - \frac{\omega_k \tau}{2} \right)^{2 \nu - 1} + \mathcal{O} \left( \tau^{4 \nu - 1} \right) , & - \frac{1}{8} < \kappa < 0 , \\
- \frac{1}{2 \tau} \frac{\cosh \left( \pi | \nu | \right) + \cos \left( 2 | \nu | \ln \frac{-\omega_0 \tau}{2} - \varphi_\nu \right) - 2 | \nu | \sin \left( 2 | \nu | \ln \frac{-\omega_0 \tau}{2} - \varphi_\nu \right)}{\cosh \left( \pi | \nu | \right) + \cos \left( 2 | \nu | \ln \frac{-\omega_0 \tau}{2} - \varphi_\nu \right)} , & \kappa < - \frac{1}{8} ,
\end{cases}
\end{equation}
where $\varphi_\nu = 2 {\rm arg} \Gamma\left(i | \nu | \right)$.

The form of these expansions has two important consequences for our discussion 
in section \ref{QF}. The leading contribution to the imaginary part of the 
matrix $\Omega$ of eq. \eqref{Omega}, (\ref{block-form})  
is proportional to the identity matrix whenever the 
index of the Bessel functions is real. However, the subleading contribution ceases being 
proportional to the couplings matrix, defined in eq. (\ref{matrixK}), 
at $\kappa = 3 / 8$ ($\nu = 1$). For smaller real values of 
$\kappa$, the subleading contribution is proportional to $K^\nu$. As a result, even in a local 
field theory, where the couplings connect neighbouring sites, the block $B$ of the matrix 
$\Omega$ does not contain only a limited number of non-vanishing elements, i.e. those which 
correspond to pairs of nodes directly coupled in the Hamiltonian. On the contrary, in general all its elements are non-vanishing.

At the same critical value of $\kappa$ another important behaviour changes. The matrix $\beta$
of eq. (\ref{betam}) behaves as $\tau^{3 - 2 \nu}$ at leading order in $\tau$ for $\kappa > 3 / 8$, and as $\tau^{2 \nu - 1}$ for $\kappa < 3 / 8$. Furthermore, the matrix $\gamma$ 
of eq. (\ref{gammam}) behaves as $\tau^{3 - 2 \nu}$ for $\kappa > 3 / 8$ and as $\tau^{- 1}$ for $\kappa < 3 / 8$. This is due to the fact that the
definition of the  matrix $\gamma$ contains two terms: the term $B^T \mathrm{Re} A^{- 1} B / 2$ is dominant for $\kappa > 3 / 8$, whereas the term $C$ is dominant for $\kappa < 3 / 8$. As a result, the matrices $\gamma$ and $\beta$ cease to tend to the same limit as $\tau \to 0^-$ for $\kappa < 3 / 8$, implying that the non-vanishing eigenvalues of the matrix $\Xi$ cease to tend to one.

The above behaviour implies that the entanglement entropy diverges at $\tau \to 0^-$ only if $\kappa > 3 / 8$. However, at this range of $\kappa$ the subleading contributions to the imaginary part of $\Omega$ are proportional to the couplings matrix, and thus such divergent terms cannot be proportional to the volume, but only to the area of the subsystem. For $\kappa < 3 / 8$, it is possible that a volume term appears in the entanglement entropy, but this cannot be divergent.

We can see this behaviour explicitly in the toy example of the two oscillators. For $\kappa > 3 / 8$, we have for $\tau \to 0^-$
\begin{align}
\gamma &= \frac{\Gamma^2 \left( \nu \right) \left( \omega_-^2 - \omega_+^2 \right)^2}{2^{5 - 2 \nu} \pi \left( \nu - 1 \right)^2 \left( \omega_-^{2 \nu} + \omega_+^{2 \nu} \right)} \left( - \tau \right)^{3 - 2 \nu} + \frac{\pi  \left( \omega_-^{2 \nu} + \omega_+^{2 \nu} \right)}{2^{1 + 2 \nu} \Gamma^2 \left( \nu \right)} \left( - \tau \right)^{2 \nu - 1} + \ldots , \\
\beta &= \frac{\Gamma^2 \left( \nu \right) \left( \omega_-^2 - \omega_+^2 \right)^2}{2^{5 - 2 \nu} \pi \left( \nu - 1 \right)^2 \left( \omega_-^{2 \nu} + \omega_+^{2 \nu} \right)} \left( - \tau \right)^{3 - 2 \nu} + \frac{\pi \left( \omega_-^{2 \nu} - \omega_+^{2 \nu} \right)}{2^{1 + 2 \nu} \Gamma^2 \left( \nu \right) \left( \omega_-^{2 \nu} + \omega_+^{2 \nu} \right)} \left( - \tau \right)^{2 \nu - 1} + \ldots
\end{align}
The above imply that
\begin{equation}
\xi = 1 - \frac{\pi^2 \left( \nu - 1 \right)^2}{\Gamma^4 \left( \nu \right) \left( \omega_-^2 - \omega_+^2 \right)} \left( 2 \omega_-^\nu \omega_+^\nu \left(  2 \omega_-^\nu \omega_+^\nu + \sqrt{2} \sqrt{\omega_-^{4 \nu} + \omega_+^{4 \nu}}\right) \right) \left( - \frac{\tau}{2} \right)^{4 \left( \nu - 1 \right)} ,
\end{equation}
which in turn implies that the entanglement entropy diverges as $\tau \to 0^-$ like
\begin{equation}
S \simeq -  4 \left(\nu - 1 \right) \ln \left( - \tau \right) . 
\end{equation}

For $\kappa < 3 / 8$ we have
\begin{align}
\gamma &= \frac{\pi}{2^{1 + 2 \nu} \Gamma^2 \left( \nu \right)} \frac{\left( \omega_-^{2 \nu} - \omega_+^{2 \nu} \right) \left( \cot \left( \pi \nu \right) + 1 \right) + 8 \omega_-^{2 \nu} \omega_+^{2 \nu}}{\left( \omega_-^{2 \nu} + \omega_+^{2 \nu} \right)} + \ldots \\
\beta &= \frac{\pi}{2^{1 + 2 \nu} \Gamma^2 \left( \nu \right)} \frac{\left( \omega_-^{2 \nu} - \omega_+^{2 \nu} \right) \left( \cot \left( \pi \nu \right) + 1 \right)}{\left( \omega_-^{2 \nu} + \omega_+^{2 \nu} \right)} + \ldots
\end{align}
The above imply that
\begin{multline}
\xi = {\left( \omega_-^{2 \nu} - \omega_+^{2 \nu} \right)^2} \bigg[2 \left( 2 \omega_-^\nu \omega_+^\nu \sin \left( \pi \nu \right) \right)^2 + \left( \omega_-^{2 \nu} - \omega_+^{2 \nu} \right)^2 \\
+ 4 \omega_-^\nu \omega_+^\nu \sin \left( \pi \nu \right) \sqrt{\left( 2 \omega_-^\nu \omega_+^\nu \sin \left( \pi \nu \right) \right)^2 + \left( \omega_-^{2 \nu} - \omega_+^{2 \nu} \right)^2}\bigg]^{-1} + \ldots
\end{multline}
As a result, the entanglement entropy tends to a finite value, namely the value that eq. \eqref{ent2osc} returns upon substitution of the above value of $\xi$. 

For imaginary arguments of the Bessel functions index, the entanglement entropy also tends to a finite value. The proof requires tedious algebra and it is beyond the scope of this work.

\acknowledgments

We would like to thank A. Karanikas and D. Katsinis for useful discussions.
The research of K. Boutivas and N. Tetradis was supported by the Hellenic Foundation for
Research and Innovation (H.F.R.I.) under the “First Call for H.F.R.I.
Research Projects to support Faculty members and Researchers and
the procurement of high-cost research equipment grant” (Project
Number: 824).

\end{document}